# Effect of electron-phonon scattering, pressure and alloying on the thermoelectric performance of TmCu$_3$Ch$_4$ (Tm=V, Nb, Ta; Ch=S, Se, Te)

Enamul Haque

EH Solid State Physics Laboratory, Longaer, Gaffargaon-2233, Mymensingh, Bangladesh.

Email: enamul.phy15@yahoo.com, enamul@mailaps.org

**Abstract**

The demand for green energy increases day by day due to environmental concern and thermoelectric (TE) materials are one of the eco-friendly energy resources. Few authors reported high TE performance in TmCu$_3$Ch$_4$, reaching the figure of merit (ZT) above 2 at 1000K, from first-principles calculations neglecting electron-phonon scattering, spin-orbit coupling effect (SOC), and energy-dependent carrier lifetime. Here, thermoelectric transport properties of TmCu$_3$Ch$_4$ are reinvestigated through considering these parameters, and significant discrepancies are found. The ZT of p-type TaCu$_3$Te$_4$ can reach ~3 at 1000K among these compounds due to its low lattice thermal conductivity ($\kappa_l$) (0.38 $W\,m^{-1}\,K^{-1}$). Interestingly, the value of $\kappa_l$ is reduced to 0.17 $W\,m^{-1}\,K^{-1}$ through 1 GPa pressure while the power factor is slightly improved due to bandgap reduction, leading to an extraordinary ZT~5.5 at 1000K. Although the substitution of Se causes a slight reduction of $\kappa_l$ to ~0.3 $W\,m^{-1}\,K^{-1}$, the power factor is reduced significantly due to the dramatic reduction of DOS near Fermi level, which leads to lower the Seebeck coefficient largely and increase electrical conductivity slightly.

## 1. Introduction

The modern development of the world increases the energy demand dramatically and the production of huge energy in a conventional way is damaging the environment severely through carbon emission. Thus, global warming becomes our great concern, and researchers all over the world are constantly searching for an alternative way, which will not cause any carbon emission, to meet the next generation of energy demand. Thermoelectric (TE) materials can generate electricity directly by using heat and have major applications, e.g. thermoelectric generator, refrigeration, etc. [1]. However, the efficiency of TE materials is very limited due to its complex nature. The TE figure of merit (ZT), a dimensionless quantity that determines the TE efficiency of a material, is defined by [2,3] $ZT = \frac{S^2\sigma}{\kappa_{tot}}T$, where S, $\sigma$, $\kappa_{tot}$, and T are the Seebeck coefficient, electrical conductivity, total thermal conductivity (electronic plus phononic), and absolute temperature and the product $S^2\sigma$ is called the power factor (PF). In general, a high Seebeck coefficient requires a large bandgap, flat bands, and large density of states (DOS) near the Fermi level. But the large bandgap and large DOS are detrimental to the electrical conductivity, as the carrier mobility is reduced by large DOS. On the other hand, when the electrical conductivity is high, the electronic thermal conductivity also becomes high. Therefore, it is preferable to reduce the lattice thermal conductivity keeping all other parameters almost the same through nanostructuring and alloying. Recent studies have been shown that pressure can enhance TE performance by reducing the lattice thermal conductivity effectively [4].

Recently, copper-based chalcogenides TmCu$_3$Ch$_4$ have been studied extensively both through theoretically and experimentally [5–13], because this family of compounds has strong potential in a broad range of applications, such as a transparent conductor and photovoltaic material [14–17]. Relatively wide bandgap, structural stability (e.g., TaCu$_3$Se$_4$ is stable up to 1328K [18]), and earth-abundant and non-toxic constituent elements have been brought great attention to the researcher to investigate its potential applicable properties. Thus, few authors reported thermoelectric properties of this family of compounds from first-principles calculations [12,13]. In Ref. [12] authors investigated TE properties of Cu$_3$MCh$_4$ (M=V, Nb, Ta; Ch=Se, Te) and reported that the ZT of p-type Cu$_3$TaTe$_4$ can reach to 2.36 at 1000K. First, in their calculations, the authors did not consider the spin-orbit coupling effect (SOC) by stating that the SOC responses weakly to the bandgap and TE properties mainly depends on the bandgap. Authors might use the default upper window energy 5.0 Ry in their calculations, which is insufficient to account SOC effect correctly [19]. Without accurate SOC calculations it is impossible to visualize the light and heavy holes (valence bands that has light mass and heavy mass, respectively) and the band degeneracy correctly [19], but authors predicted that high band degeneracy is the main source of the high TE performance of this family. Authors found that valence bands at Γ-point are three-fold degenerate bands but it is not visible from the band structure without the SOC effect [12]. Furthermore, authors considered DP theory to calculate the constant carrier lifetime neglecting electron-phonon scattering, and thus, reported a large power factor of 12.53 mWm$^{-1}$K$^{-2}$ at 300K [12]. Unfortunately, in evaluating the PF, the authors made a serious mistake in Table 4 [12]. Besides, authors also made a serious mistake in their elastic constants and all other related parameter calculations, such as Debye temperature, lattice thermal conductivity, and hence, thermoelectric figure of merit [12]. It is not clear why authors did not verify their calculated elastic constants, as several articles reported the elastic constants of this family of compounds [20–23]. On the other hand, in Ref. [13] authors reported TE properties of VCu$_3$X$_4$ (X=S, Se, Te) from first-principles calculations neglecting SOC effect and energy-dependent carrier lifetime. However, in this article, the authors calculated lattice thermal conductivity accurately and reported large discrepancies of their TE properties compared to that reported in Ref. [12]. From the above discussion, it is clear that an accurate study needs to be reported to assess the actual TE performance of this family of compounds and remove the ambiguity.

Here, thermoelectric properties of TmCu$_3$Ch$_4$ are reported from first-principles calculations considering SOC effect, and electron-phonon scattering. Furthermore, the effect of pressure and alloying through Se on the TE performance are also explored and found that 1 GPa pressure can reduce lattice thermal conductivity of TaCu$_3$Te$_4$ dramatically, leading to an extraordinary ZT~5.5 at 1000K for the p-type carrier.

## 2. Computational details

First-principles calculations were performed in Quantum Espresso [24] and Wien2k [25] codes. First, structural relaxations were performed by minimizing energy and internal forces in Quantum espresso, with generalized gradient approximation (GGA) [26] of PBE functional setting [27], 6 × 6 × 6 k-point and cutoff energy for wave functions above 45 Ry, projector augmented plane wave (PAW) pseudopotential basis (provided in the PSlibrary1.0 [28]), energy convergence threshold $10^{-10}$ $Ry$, force convergence threshold $10^{-5}$ $Au/Ry$, and Marzari-Vanderbilt smearing [29] of width 0.001 Ry. By using the fully relaxed structures, the electron-phonon (e-ph) dynamical matrix was calculated by using 8 × 8 × 8 k-point, 444 q-point, energy convergence threshold $10^{-12}$ $Ry$, and keeping other setting same. The resulted e-ph dynamical matrix were then feed into EPA code [30] to calculate average e-ph matrix. The carrier lifetime was calculated by using the average e-ph matrix in slightly modified BoltzTraP code [31,30]. In this modified code, the carrier life time is calculated through the equation [30]

$$\tau^{-1}(\epsilon, \mu, T) = \frac{2\pi\Omega}{g_s \hbar} \sum_v \{g_v^2(\epsilon, \epsilon + \bar{\omega}_v)[n(\bar{\omega}_v, T) + f(\epsilon + \bar{\omega}_v, \mu, T)] \rho(\epsilon + \bar{\omega}_v)$$
$$+ g_v^2(\epsilon, \epsilon - \bar{\omega}_v)[n(\bar{\omega}_v, T) + 1 - f(\epsilon - \bar{\omega}_v, \mu, T)]\rho(\epsilon - \bar{\omega}_v)\} \ldots \ldots (1)$$

where $\Omega$ is the volume of the primitive cell, $\hbar$ is the reduced Planck's constant, $v$ the index of phonon modes, and $\bar{\omega}_v$ the averaged frequency of phonon modes. The $g_v^2$ is the averaged e-ph matrix, $n(\bar{\omega}_v, T)$ is the Bose-Einstein distribution function, $f(\epsilon + \bar{\omega}_v, \mu, T)$ the Fermi-Dirac distribution function, $g_s = 2$ the degeneracy of spin, $\epsilon$ the energy of electrons, and $\rho$ is the electronic DOS per unit volume and energy. Please see Ref. [30] for the details of the equation. The other parameters (such as energy eigenvalues) required for BoltzTraP calculations were obtained from Wien2k, a full-potential linearized augmented plane wave (FP-LAPW) based code [25], by using PBE functional with mBJ potential [32,33] including SOC effect, 40 × 40 × 40 k-point mesh, and kinetic energy cutoff RKmax=7.0 (the product of the smallest atomic sphere radius and the largest K-vector (i.e., the plane wave expansion of the wave function), thus, it determines the size of the basis set). The smaller value of the Rmt leads to a larger number of plane waves (PWs) expansion and requires a lower value of RKmax. After extensive trials, the values of the muffin tin sphere were set and are listed in Table S1 (see ESI).

The lattice thermal conductivity ($\kappa_l$) and related parameters were calculated by using finite displacement methods as implemented in Phono3py [34]. For this, the second and third-order interatomic force constants (IFCs) were calculated by creating 222 supercells in Quantum Espresso with 2 × 2 × 2 k-point and keeping other settings the same. The $\kappa_l$ was then calculated by solving the phonon Boltzmann transport equation with 121212 q-point. Please see the supplementary information (ESI) for further details.

## 3. Results and discussion

TmCu$_3$Ch$_4$ compounds form a simple cubic structure of space group $P\bar{4}3m$ (#215) with one formula unit (Z=1) [7]. The Tm and Cu cations are positioned at the corner and center of the edge, respectively, while anions (Ch) are located at the tetrahedral positions. The fully relaxed lattice parameters of TmCu$_3$Ch$_4$ are listed in Table I and compared with the experimental values.

Table I. Comparison of fully optimized lattice parameters of TmCu$_3$Ch$_4$ (Tm=V, Nb, Ta; Ch=S, Se, Te), TaCu$_3$Te$_4$ under 1 GPa pressure, and TaCu$_3$Te$_3$Se by using PBE functional with the available experimental values.

| Compounds | Calc. lattice parameters (Å) | Exp. |
|---|---|---|
| VCu$_3$S$_4$ | 5.438 | 5.393 [7] |
| VCu$_3$Se$_4$ | 5.647 | 5.563 [8] |
| VCu$_3$Te$_4$ | 5.924 | - |
| NbCu$_3$S$_4$ | 5.543 | 5.500 [35] |
| NbCu$_3$Se$_4$ | 5.725 | 5.638 [35] |
| NbCu$_3$Te$_4$ | 6.003 | 5.921 [36] |
| TaCu$_3$S$_4$ | 5.558 | 5.520 [35] |
| TaCu$_3$Se$_4$ | 5.732 | 5.670 [35] |
| TaCu$_3$Te$_4$ | 6.009 | 5.928 [37] |
| TaCu$_3$Te$_4$ (1 GPa) | 5.948 | - |
| TaCu$_3$Te$_3$Se | 5.947 α=89.06° | - |

The optimized lattice parameters fairly agree with the experiment, the results overestimate the experiment by 1%, which is a known limitation of PBE functional. Note that the deviations in the case of the heavier compound are slightly higher, which might be due to the exclusion of the SOC effect in the structural relaxations. The computed values of the lattice parameters are almost the same as that of reported in Ref. [12], thus all other parameters should be consistent where these are relevant. Interestingly, 1 GPa pressure reduces the lattice parameters of about 1% of TaCu$_3$Te$_4$. When one Te is replaced by Se and symmetry is applied, the symmetry is reduced to $R\bar{3}m$ (#160), a rhombohedral unit cell with three formula unit (Z=3). The lattice length of the cell is almost same to that of TaCu$_3$Te$_4$ under 1 GPa, but the angle is slightly reduced.

## 3.1. Lattice dynamics

The calculated elastic constants and related parameters are listed in Table II. All these compounds are elastically stable, based on the stability criteria described Ref. [38]. These values fairly agree with the computational results reported in Refs [20–23]. and materials projects [23]. But in Ref. [12] authors reported different values of the elastic constants.

Table II. Calculated elastic constants ($c_{ij}$), (Hill average [39]) bulk modulus (B), and shear modulus (G) (in GPa), Pugh's ratio (B/G [40]), Poisson ratio ($v$), longitudinal sound velocity ($v_l$), transverse sound velocity ($v_t$), average sound velocity ($v_a$) (in kms$^{-1}$), and Debye temperature ($\theta_D$ in K [41]).

| Compounds | $c_{11}$ | $c_{12}$ | $c_{44}$ | B | G | B/G | $v$ | $v_l$ | $v_t$ | $v_a$ | $\theta_D$ |
|---|---|---|---|---|---|---|---|---|---|---|---|
| VCu$_3$S$_4$ | 93 | 18 | 20 | 43 | 26 | 1.65 | 0.25 | 4.51 | 2.59 | 2.88 | 316 |
| VCu$_3$Se$_4$ | 69 | 16 | 18 | 34 | 21 | 1.62 | 0.24 | 3.47 | 2.019 | 2.24 | 236 |
| VCu$_3$Te$_4$ | 57 | 19 | 18 | 31 | 18 | 1.72 | 0.25 | 3.06 | 1.755 | 1.95 | 196 |
| NbCu$_3$S$_4$ | 95 | 16 | 18 | 42 | 25 | 1.68 | 0.25 | 4.35 | 2.506 | 2.78 | 299 |
| NbCu$_3$Se$_4$ | 75 | 14 | 17 | 34 | 22 | 1.54 | 0.24 | 3.48 | 2.037 | 2.26 | 235 |
| NbCu$_3$Te$_4$ | 59 | 15 | 19 | 30 | 20 | 1.5 | 0.22 | 3.06 | 1.832 | 2.02 | 201 |
| TaCu$_3$S$_4$ | 91 | 14 | 18 | 40 | 24 | 1.67 | 0.24 | 3.88 | 2.253 | 2.5 | 268 |
| TaCu$_3$Se$_4$ | 71 | 13 | 17 | 33 | 21 | 1.57 | 0.24 | 3.19 | 1.888 | 2.09 | 217 |
| TaCu$_3$Te$_4$ | 56 | 15 | 19 | 29 | 20 | 1.45 | 0.21 | 2.87 | 1.725 | 1.90 | 189 |
| TaCu$_3$Te$_4$ (1 GPa) | 62 | 18 | 21 | 32 | 21 | 1.52 | 0.23 | 2.98 | 1.768 | 1.95 | 196 |
| TaCu$_3$Te$_3$Se | 61 | 17 | 19 | 32 | 21 | 1.52 | 0.23 | 3.00 | 1.779 | 1.97 | 197 |

For example, the authors reported the values of $c_{11}$, $c_{12}$, and $c_{44}$ of TaCu$_3$Te$_4$ to be 93.81, 50.82, and 53.69 GPa, respectively [12]. Although authors used the same code (as the Materials project did [23]), their values unexpectedly differ from the values reported in Refs. [20,22] and Materials projects [23]. Authors might make a serious mistake in their calculations.

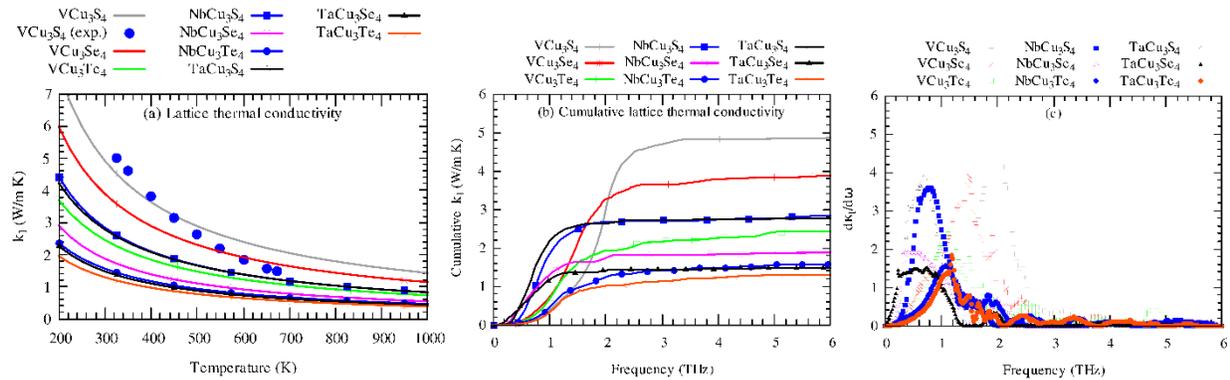

Fig. 1. Lattice thermal conductivity (a), cumulative lattice thermal conductivity (b), and it's derivative (c) of pristine TmCu$_3$Ch$_4$ (Tm=V, Nb, Ta; Ch=S, Se, Te). The blue circles represent the values of experimental lattice thermal conductivity taken from Ref. [42].

As the Ref. [12] reported wrong elastic constants, all other parameters are seemed to be wrong. In general, the semiconducting materials are brittle, which is also valid for this family of compounds

(from Pugh's ratio), but Ref. [12] reported to be ductile. The sound velocity propagates less likely in the heavier compound and thus, the Debye temperature decreases with the change from V to Ta and S to Te. As usual, the lattice thermal ($\kappa_l$) of heavier compounds is relatively low than that of lighter compounds. The Debye temperature shows a similar trend because $\kappa_l$ is directly related to the Debye temperature.

Fig. 1 shows the computed lattice thermal conductivity and related parameters. The computed value of $\kappa_l$ of VCu$_3$S$_4$ (blue circles in Fig. 1(a)) fairly agrees with the experiment [42] at low temperature, but it rapidly diverges with temperature. Above 400 K, the $\kappa_l$ overestimates the value, about 30% at 650 K. At high temperature, there might be some complex scattering mechanisms involving these materials and pBTE fails to account for these scatterings correctly. The calculated $\kappa_l$ also agrees well with the theoretical values reported in Ref. [13]. But the scenario is completely different in Ref. [12], e.g. the referred article reported the $\kappa_l$ of TaCu$_3$Te$_4$ to be 5.18 at 300K, but it is found to be 1.29 Wm$^{-1}$ K$^{-1}$. As the authors used the wrong Debye temperature and other related parameters obtained from the wrong values of elastic constants, such types of results are expected. The $\kappa_l$ goes down with both M and Ch groups. This trend is consistent with the calculated Debye temperature. Acoustic phonons contribute to the heat conductions, while optical phonons have almost negligible contributions in all cases, which can be seen from the cumulative lattice thermal conductivity (Fig. 1 (b)) and its derivative ((Fig. 1 (c)). From the partial phonon density of states (Fig. S1, see ESI), The Cu has dominated contributions to the acoustic phonons in S-based compounds, while Se/Te dominantly contributes to the acoustic phonons in all other cases. Thus, Cu induced phonons are responsible for the relatively high lattice thermal conductivity of S-based compounds. Phonons dispersions of all compounds are well behaved suggesting that the compounds under consideration are dynamically stable (Fig. S1, see ESI).

The phonon lifetime does not show a similar trend of increase of scattering in heavier compounds (Fig. S2 and Fig. S3, see ESI). On the other side, the mode Gruneisen parameter ($\gamma$) goes to negative in Nb/Ta and Se/Te based compounds. The heaviest compound has the largest value of mode Gruneisen parameter, especially for the acoustic phonons (Fig. S4, see ESI). The $\gamma$ indicates the anharmonicity and the larger value of it means intense phonon scattering and vice versa. Therefore, the trend of $\gamma$ suggests that the heavier atoms can induce intense phonon scattering, which is valid in general for other materials. Thus, TaCu$_3$Te$_4$ has the lowest lattice thermal conductivity among these compounds (1.29 Wm$^{-1}$ K$^{-1}$ at 300K) and are expected to be the most efficient thermoelectric material at high temperature. For this reason and because of their isophononic structure, the effect of pressure and alloying on the thermoelectric performance of TaCu$_3$Te$_4$ will only be studied.

### 3.2. Electronic structure

TmCu$_3$Ch$_4$ compounds are semiconductors and the computed values of the bandgap of these compounds by mBJ potential including the SOC effect are listed in Table III. The maxima of the valence band (VBM) and the minima of the conduction band (CBM) lie at two different K-point, suggesting the indirect nature of the bandgap ((Fig. S5, see ESI)). The computed bandgap of VCu$_3$S$_4$ underestimates the experimental value by 10% [17]. The values of the bandgap are consistent with the available theoretical values. The gaps widen up with the Tm group and narrow with Ch group, which is also consistent with the Ref. [15].

Table III. Computed bandgap, effective mass of the heavy hole, light hole, and electrons (in the unit of $m_0$)

| Compounds | Calc. bandgap (eV) (mBJ+SOC) | PBEsol+U (Ref. [15]) (eV) | $m_h^*$ (heavy hole) | $m_h^*$ (light hole) | $m_e^*$ (electron) |
| --- | --- | --- | --- | --- | --- |
| VCu$_3$S$_4$ | 1.17 (1.3 exp. [17]) | 1.13 | 1.77 | 1.75 | 3.41 |
| VCu$_3$Se$_4$ | 1.12 | 0.87 | 2.44 | 2.39 | 3.66 |
| VCu$_3$Te$_4$ | 0.67 | 0.53 | 0.95 | 0.97 | 5.5 |
| NbCu$_3$S$_4$ | 1.73 | 1.82 | 1.63 | 1.60 | 1.72 |
| NbCu$_3$Se$_4$ | 1.50 (2.14 exp. [18]) | 1.45 | 1.75 | 1.72 | 1.52 |
| NbCu$_3$Te$_4$ | 1.02 | 0.92 | 0.87 | 0.83 | 1.68 |
| TaCu$_3$S$_4$ | 2.04 (2.70 exp. [14]) | 2.10 | 1.81 | 1.78 | 1.92 |
| TaCu$_3$Se$_4$ | 1.78 (2.35 exp. [14]) | 1.71 | 1.85 | 1.89 | 1.93 |
| TaCu$_3$Te$_4$ | 1.25 | 1.11 | 0.95 | 0.94 | 1.86 |
| TaCu$_3$Te$_4$ (1 GPa) | 1.20 | - | 0.83 | 0.77 | 1.73 |
| TaCu$_3$Te$_3$Se | 1.23 | - | 0.54 | 0.68 | 2.4 |

Both the lowest conduction bands are and the highest valence bands are four-fold degenerate bands, but the Ref. [12] reported that the lowest conduction bands are two-fold and the highest valence bands are three-fold degenerate bands, which is not correct as it considered a low value of upper window energy, 5.0 Ry, in SOC calculations. Such high band degeneracy arises from the four three-fold rotational symmetries. Both CBM and VBM have two extrema, leading to the number of degeneracy eight ($N_v$=8).

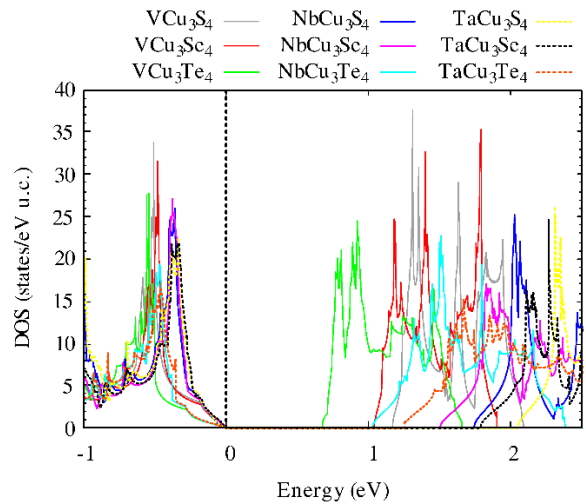

Fig. 2. A comparison of the total density of states of pristine TmCu$_3$Ch$_4$ (Tm=V, Nb, Ta; Ch=S, Se, Te). The dashed line at zero energy represents the Fermi level. Please see the supplementary material for their corresponding band structures (as shown in Fig. S).

Cations M-d and Cu-d orbitals mainly formulate the bandgap with a small contribution from the Ch-p states. The computed bandgap also suggests the relative weak dependency on the Ch-p compared to M-d states. The electron effective mass is heavier than that of holes and the difference of masses between light and heavy holes are small. This suggests that the transport properties of the p-type carrier would be much better compared to the n-type carrier. The calculated hole effective mass is much higher than that reported in Ref. [12] (reported to be ~0.3 $m_0$ for $Cu_3MCh_4$ (M=V, Nb, Ta; Ch=Se, Te)), which is unexpected, but consistent with Ref. [13]. Fig. 2 shows the total density of states (DOS) of $TmCu_3Ch_4$. The DOS near the Fermi level of S/Se based compounds is higher than that of Te based compounds, suggesting that Te based compounds would have better TE performance compared to others.

### 3.3. Thermoelectric properties

The computed energy-dependent carrier lifetime falls sharply on the band edge (see Fig. S6 of ESI), which follows the relation described in Ref. [30].

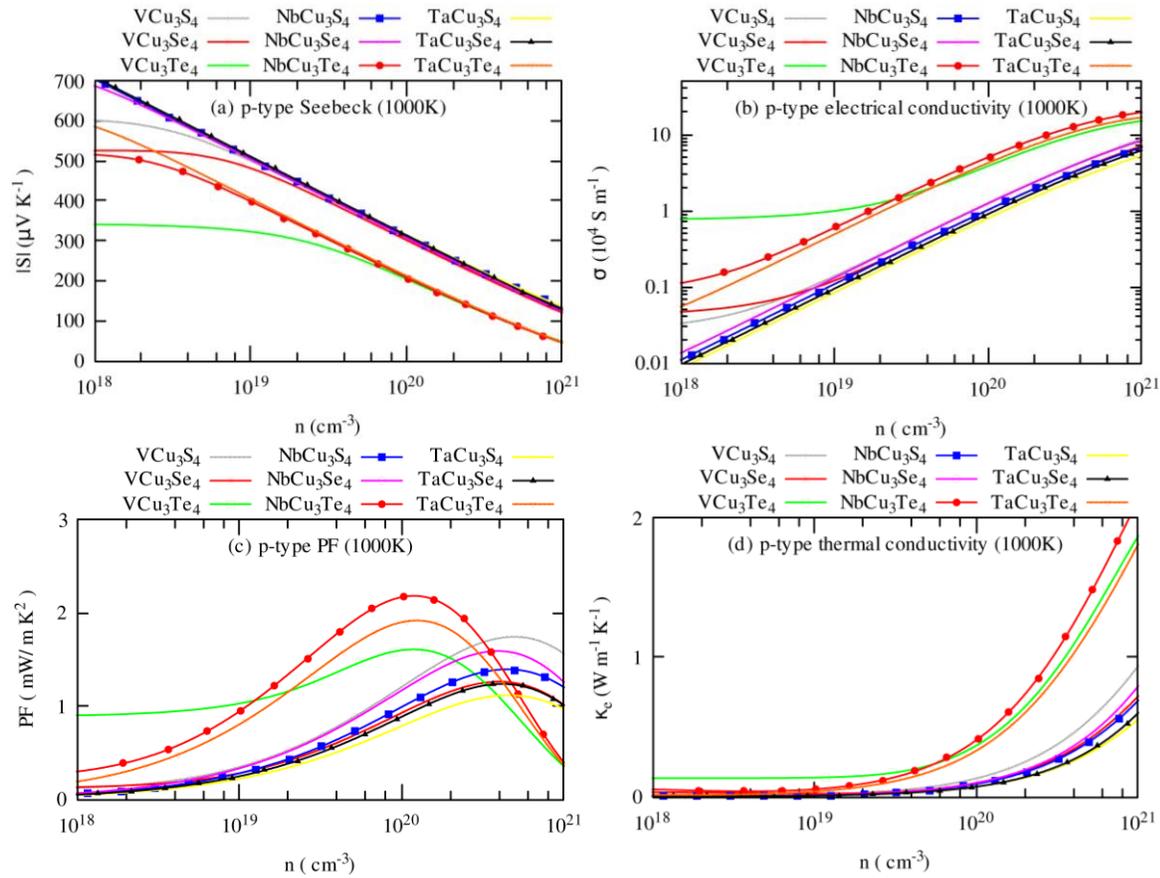

Fig. 3. Carrier concentration-dependent transport coefficients of p-type TmCu3Ch4 (Tm=V, Nb, Ta; Ch=S, Se, Te) at 1000K: (a) Seebeck coefficient, (b) electrical conductivity, (c) power factor (PF), and (a) electronic part of the thermal conductivity ($\kappa_e$).

Fig. 3 shows the carrier concentration dependence transport coefficients for the p-type carrier at 1000K (please see Fig. S7 and Fig. S8 for transport coefficients n- and p-type carriers at 300 K in ESI). Seebeck coefficients sharply fall with carrier concentrations while the electrical conductivity and electronic part of the thermal conductivity exhibit the opposite trends. The Seebeck coefficients are very large due to wide bandgap, heavy effective mass, and high density of states near Fermi level. As the bandgap of Te compounds is smaller, the DOS near the Fermi level is lower and effective mass is lighter compared to other compounds, the Seebeck coefficients of Te-based compounds are lower. On the other side, the electrical conductivities of Te-based compounds are much higher than that of S/Se based one. The electronic part of the thermal conductivity ($\kappa_e$) of these compounds also shows the same trend. The S of n-type carriers is larger than that of p-type carriers due to the heavier effective mass of electrons. But both the electrical conductivity and electronic part of the thermal conductivity of n-type carrier are lower because the heavy effective mass of carrier is non-conducive and reduce the mobility of the carrier significantly.

The large value of S and reasonable value of the electrical conductivity of p-type carriers leads to a larger power factor (PF) as shown in Fig. 3(c). The relatively lower value of electrical conductivity of n-type carriers leads to almost two times smaller power factor, although the Seebeck coefficients are larger. This suggests the importance of reasonable electrical conductivity of thermoelectric materials. The power factor reported in the Ref. [12] is much larger than that of the present one, e.g. the PF reported to be 12.53 $mW\,m^{-1}\,K^{-2}$ for the p-type carrier at 300K, but the present study reveals that it should be ~4.5 $mW\,m^{-1}\,K^{-2}$. The use of wrong effective mass and elastic constants in evaluating the carrier lifetime might be responsible for a big difference. On the other side, the authors of the article made a serious mistake in evaluating n-type PF (see Table 4 of Ref. [12]).

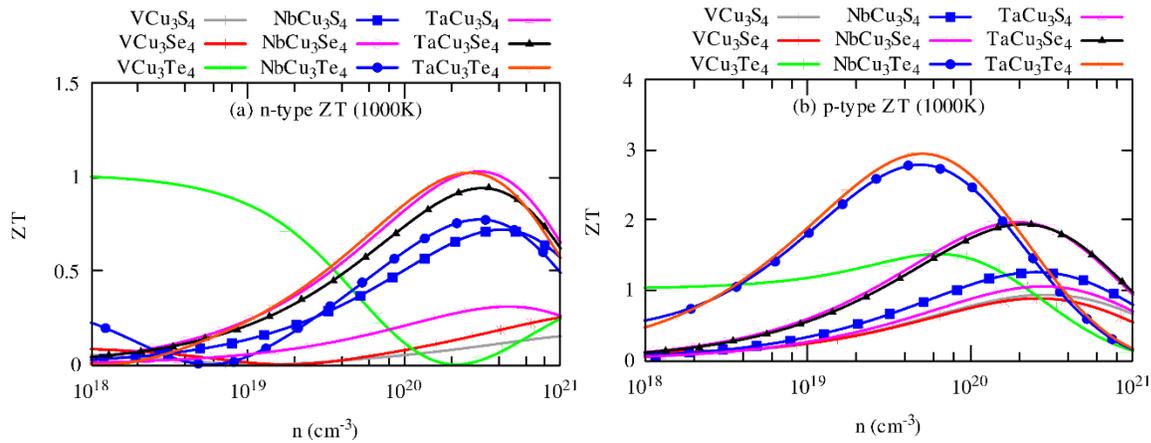

Fig. 4. Carrier concentration dependency of thermoelectric figure of merit (ZT) of TmCu$_3$Ch$_4$ (Tm=V, Nb, Ta; Ch=S, Se, Te) for (a) n-type and (b) p-type carrier at 1000K.

Large PF and low lattice thermal conductivity of Te-based compounds (except VCu$_3$Te$_4$) leads to the highest ZT among these compounds, as shown in Fig. 4 (see Fig. S9 for ZT at 300K and Fig. S10 for transport coefficients of n-type carriers at 1000K). The bandgap of VCu$_3$Te$_4$ is too small for high-temperature applications because of the carrier exciton at high temperatures. Thus, the ZT of VCu$_3$Te$_4$ is much smaller than that of NbCu$_3$Te$_4$ and TaCu$_3$Te$_4$. Thus, V-based copper chalcogenides are not suitable for TE applications, although Ref. reported a high ZT~1.6 at 1000K for p-type carriers of VCu$_3$Se$_4$/Te$_4$ (Authors found a high ZT because of the use energy independent carrier lifetime). The ZT of p-type carrier is slightly larger than that reported in Ref. [12] for certain cases, as the authors used the wrong lattice thermal conductivity in the evaluation of TE performance. The maximum ZT of n-type carrier is ~1 at 1000K, suggesting the n-type TmCu$_3$Ch$_4$ compounds are not suitable for TE applications unless ZT is improved by alloying or nanostructuring. The low-temperature ZT is also small for both n- and p-type carriers. Therefore, TmCu$_3$Ch$_4$ compounds are not suitable for low-temperature TE applications. The low-temperature TE performance might be improved by reducing the bandgap, as the best known low-temperature TE materials (such as Bi$_2$Te$_3$ and its alloys [43,44]) have a narrower bandgap compared to that of TmCu$_3$Ch$_4$. From these points, the next section will be devoted to addressing this issue.

### 3.4. Effect of pressure and alloying

Fig. 5 shows the phonon dispersions of TaCu$_3$Te$_4$ at 0 and 1 GPa pressure and Se-substituted TaCu$_3$Te$_4$ are shown in Fig. 5 (optical phonons of higher frequencies are omitted here for clarity). The left panel shows the total phonon density of states. Pressure causes acoustic phonon softening significantly and when 2 GPa pressure is applied, acoustic phonons frequency becomes negative suggesting structural instability at 2 GPa. For this reason, the present study only focuses on the 1 GPa pressure. The optical phonons frequencies are also changed.

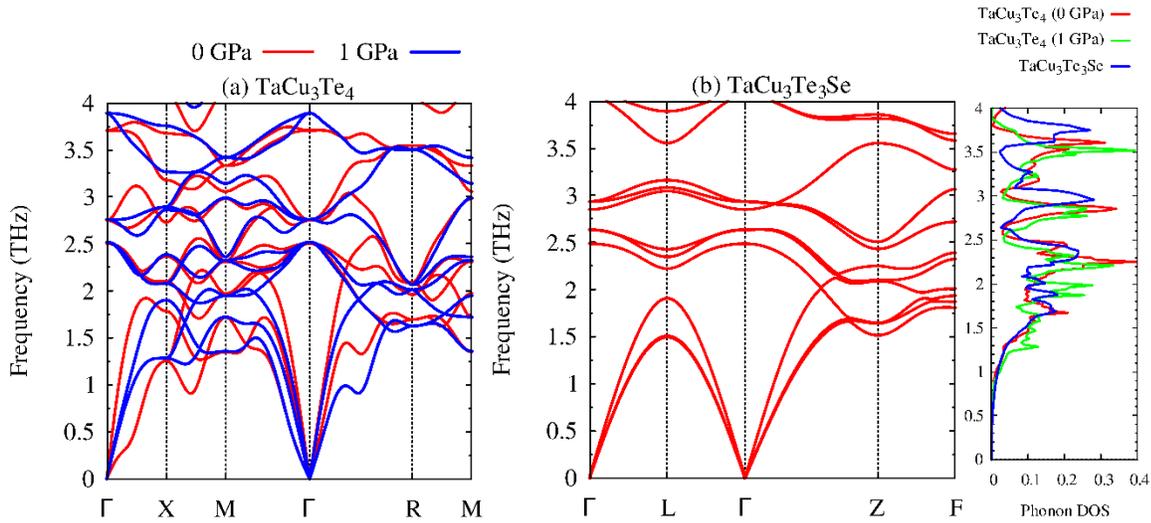

Fig. 5. Visualization of the effect of pressure and alloying through Se on the phonon dispersion and density of states of TaCu$_3$Te$_4$. The right panel shows the total phonon density of states of TaCu$_3$Te$_4$ at ambient conditions, under 1 GPa pressure, and its alloys.

The acoustic phonon DOS increases significantly in the lower frequency region. But the substitution of Se does not cause phonon softening, rather it opens several gaps among optical phonons, and between acoustic phonons and optical phonons along Γ to Γ point, which is conducive for heat. Besides, the acoustic phonons DOS remains almost unchanged, suggesting the phonon scattering does not improve effectively. As Se is a lighter element compared to Te, this type of change is expected. The enhancement of phonon scattering leads to a dramatic reduction of lattice thermal conductivity as shown in Fig. 6(a). Although pressure softens the acoustic phonons, the highest peak of cumulative lattice thermal conductivity and its derivative remains almost at the same energy.

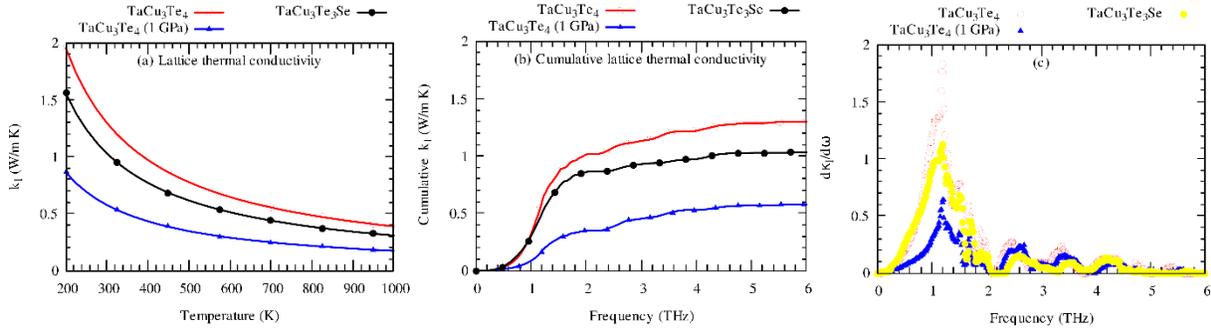

Fig. 6. Comparison of lattice thermal conductivity and related parameters of TaCu$_3$Te$_4$ under 1 GPa pressure and alloying through Se.

As Se-substitution does not enhance phonon scattering efficiently, the lattice thermal conductivity is reduced by only a few percent. The computed lifetime suggests the same features (see Fig. S3 in ESI). Moreover, the mode Gruneisen parameter is also increased higher by pressure compared to that by Se-substitution. In general, a lighter element is less effective to reduce the lattice thermal conductivity. This dramatic reduction of lattice thermal conductivity would cause a substantial improvement in TE performance of TaCu$_3$Te$_4$, but the major concern is the electronic structure, if it changes too much, TE performance might be hampered instead of improvement.

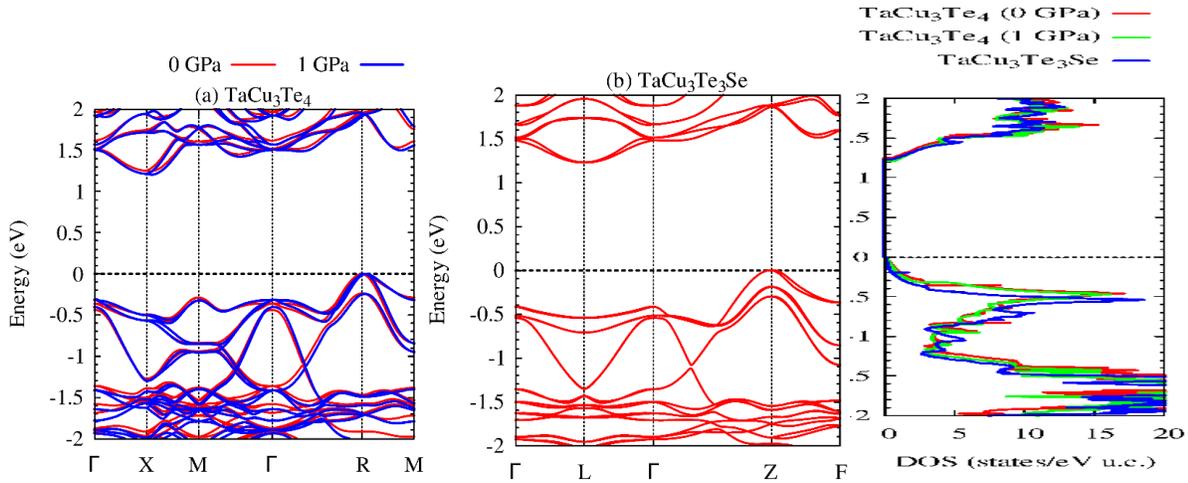

Fig. 7. Demonstration of the change of electronic structure of TaCu$_3$Te$_4$ under 1 GPa pressure and alloying through Se. The right panel shows the total electronic density of states of TaCu$_3$Te$_4$ at ambient conditions, under 1 GPa pressure, and its alloys. The grey colored dashed line at zero energy represents the Fermi level.

Fortunately, it doesn't change significantly, as shown in Fig. 7, and thank to pressure for such type changes. The bandgap is reduced slightly but the band degeneracy, shape of the CBM, and VBM remain almost unchanged. Unfortunately, the Se-substitution reduces the degeneracy of the lowest valence bands to two-fold from four-fold degenerate bands, although it does not changes the conduction bands significantly. Moreover, the density of states near the Fermi level is lowered significantly, which would be detrimental for high TE performance. Furthermore, the effective mass of holes dramatically lightens by Se-substitution. Interestingly, the DOS near the Fermi level remains almost unchanged under 1 GPa pressure while the effective mass of holes is reduced slightly (see Fig. S11 for the projected density of states in ESI).

Fig. 8 demonstrates the effect of pressure and Se-substitution on the transport coefficients of p-type carriers at 1000K (see Fig. S12, Fig. S13, and Fig. S14 for transport coefficients of n-and p-type carriers at 300K, and n-type carriers at 1000K, respectively in ESI). As the 1 GPa pressure reduces the bandgap and effective mass slightly, the Seebeck coefficient is reduced slightly while the electrical conductivity is increased. Interestingly, the change in electrical conductivity is predominant than that of the Seebeck coefficient. Thus, the PF is improved by ~15% under 1 GPa pressure, due to such impressive change (as shown in Fig. 8 (c)).

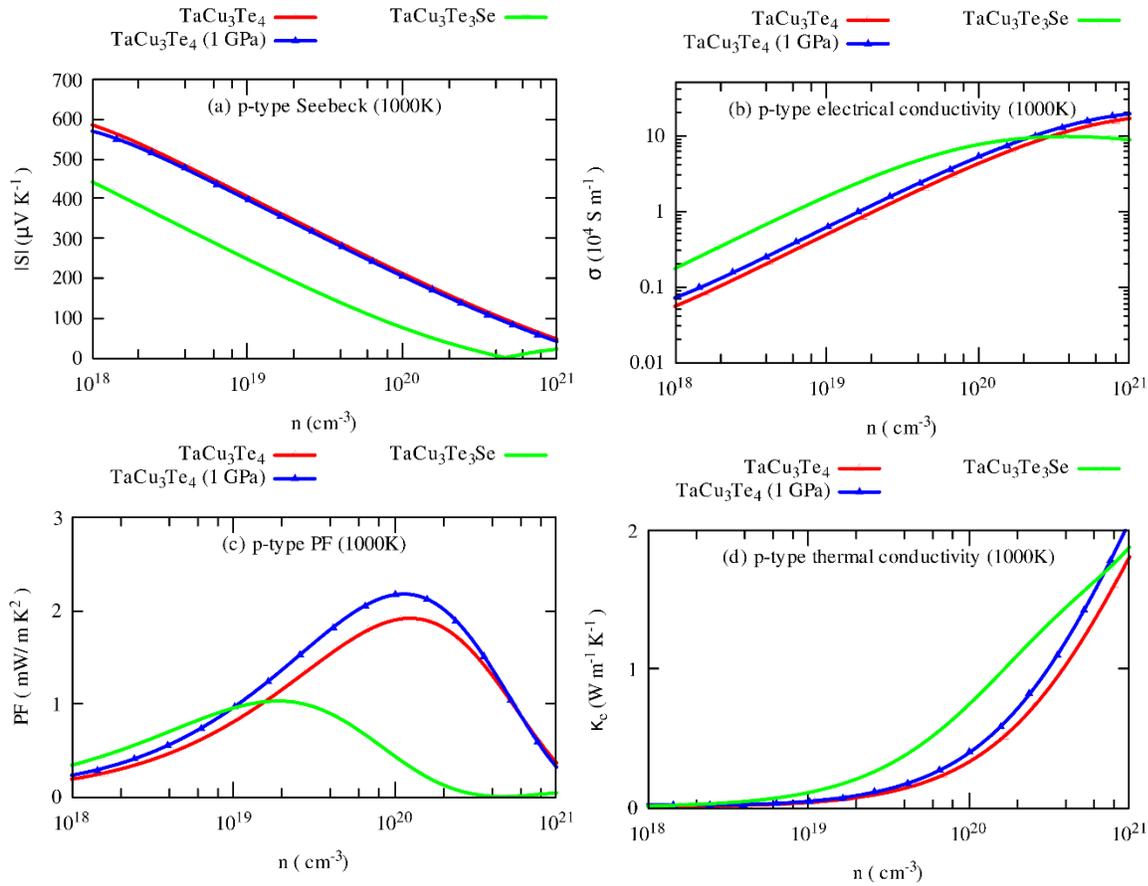

Fig. 8. Carrier concentration-dependent transport coefficients (at 1000K) of p-type $TaCu_3Te_4$ at ambient condition, under 1 GPa pressure, and alloyed through Se.

Table IV. A comparison of computed transport coefficients and other related parameters (optimum carrier concentration ($n_{opt}$ ($10^{20}$ $cm^{-3}$), Seebeck coefficient (S (μV K$^{-1}$)), electrical conductivity ($\sigma$ ($10^4$ $S$ $m^{-1}$)), power factor (PF ($mW$ $m^{-1}K^{-2}$)), total thermal conductivity ($\kappa_{tot}$ ($Wm^{-1}K^{-1}$)), and thermoelectric figure of merit (ZT)) for n- and p-type carriers of all cases at 300K and 1000K.

| Compounds | | T | $n_{opt}$ | S | $\sigma$ | PF | $\kappa_{tot}$ | ZT |
|---|---|---|---|---|---|---|---|---|
| VCu$_3$S$_4$ | n | 300 | 2.034 | 198.53 | 1.778 | 0.700 | 4.965 | 0.042 |
| | | 1000 | 19.83 | 197.04 | 0.635 | 0.246 | 1.505 | 0.163 |
| | p | 300 | 0.282 | 190.38 | 6.887 | 2.496 | 5.216 | 0.143 |
| | | 1000 | 2.936 | 223.02 | 3.340 | 1.661 | 1.780 | 0.933 |
| VCu$_3$Se$_4$ | n | 300 | 2.193 | 191.86 | 1.653 | 0.608 | 3.946 | 0.046 |
| | | 1000 | 20.62 | 197.55 | 0.890 | 0.347 | 1.265 | 0.274 |
| | p | 300 | 0.499 | 187.81 | 6.180 | 2.180 | 4.158 | 0.157 |
| | | 1000 | 2.624 | 222.32 | 2.462 | 1.216 | 1.381 | 0.880 |
| VCu$_3$Te$_4$ | n | 300 | 1.838 | 192.17 | 1.053 | 0.389 | 2.496 | 0.046 |
| | | 1000 | 0.020 | 346.38 | 0.712 | 0.855 | 0.869 | 0.983 |
| | p | 300 | 0.137 | 204.21 | 8.037 | 3.351 | 2.770 | 0.363 |
| | | 1000 | 0.630 | 237.73 | 2.660 | 1.503 | 0.994 | 1.511 |
| NbCu$_3$S$_4$ | n | 300 | 0.939 | 194.33 | 5.668 | 2.140 | 3.043 | 0.211 |
| | | 1000 | 4.162 | 213.68 | 1.417 | 0.647 | 0.901 | 0.717 |
| | p | 300 | 0.232 | 201.99 | 7.004 | 2.857 | 3.110 | 0.275 |
| | | 1000 | 2.546 | 235.25 | 2.377 | 1.315 | 1.048 | 1.255 |
| NbCu$_3$Se$_4$ | n | 300 | 0.724 | 195.53 | 4.515 | 1.726 | 2.053 | 0.252 |
| | | 1000 | 3.000 | 220.65 | 1.326 | 0.645 | 0.628 | 1.027 |
| | p | 300 | 0.368 | 208.90 | 8.196 | 3.576 | 2.207 | 0.485 |
| | | 1000 | 1.921 | 252.68 | 2.261 | 1.443 | 0.734 | 1.966 |
| NbCu$_3$Te$_4$ | n | 300 | 0.773 | 189.58 | 3.014 | 1.083 | 1.687 | 0.192 |
| | | 1000 | 3.055 | 218.08 | 0.891 | 0.424 | 0.549 | 0.772 |
| | p | 300 | 0.101 | 223.20 | 9.688 | 4.827 | 1.930 | 0.750 |
| | | 1000 | 0.495 | 265.50 | 2.692 | 1.898 | 0.681 | 2.784 |
| TaCu$_3$S$_4$ | n | 300 | 1.129 | 183.70 | 2.566 | 0.866 | 2.876 | 0.090 |
| | | 1000 | 4.619 | 207.35 | 0.622 | 0.267 | 0.868 | 0.308 |
| | p | 300 | 0.267 | 198.15 | 6.388 | 2.508 | 3.028 | 0.248 |
| | | 1000 | 2.793 | 230.74 | 2.003 | 1.066 | 1.014 | 1.051 |
| TaCu$_3$Se$_4$ | n | 300 | 0.938 | 189.50 | 3.189 | 1.145 | 1.615 | 0.212 |
| | | 1000 | 3.155 | 221.52 | 0.966 | 0.474 | 0.505 | 0.939 |
| | p | 300 | 0.439 | 207.67 | 6.903 | 2.977 | 1.761 | 0.507 |
| | | 1000 | 2.106 | 251.38 | 1.789 | 1.131 | 0.584 | 1.936 |
| TaCu$_3$Te$_4$ | n | 300 | 0.842 | 191.25 | 3.016 | 1.103 | 1.427 | 0.232 |
| | | 1000 | 2.621 | 225.90 | 0.923 | 0.471 | 0.462 | 1.021 |
| | p | 300 | 0.103 | 225.89 | 8.411 | 4.291 | 1.618 | 0.794 |
| | | 1000 | 0.512 | 267.19 | 2.327 | 1.661 | 0.565 | 2.945 |
| TaCu$_3$Te$_4$ (1 GPa) | n | 300 | 0.616 | 214.44 | 2.998 | 1.378 | 0.704 | 0.587 |
| | | 1000 | 1.744 | 256.90 | 0.887 | 0.586 | 0.243 | 2.406 |
| | p | 300 | 0.061 | 258.97 | 6.551 | 4.393 | 0.822 | 1.603 |
| | | 1000 | 0.289 | 307.67 | 1.686 | 1.596 | 0.297 | 5.368 |
| TaCu$_3$Te$_3$Se | n | 300 | 0.282 | 190.42 | 1.986 | 0.720 | 1.112 | 0.194 |
| | | 1000 | 0.956 | 215.62 | 0.604 | 0.280 | 0.352 | 0.796 |
| | p | 300 | 0.017 | 217.20 | 5.497 | 2.593 | 1.243 | 0.625 |
| | | 1000 | 0.097 | 249.78 | 1.515 | 0.945 | 0.414 | 2.282 |

As usual, the electronic part of the thermal conductivity is also increased slightly and shows the same trend as that of electrical conductivity under 1 GPa. As the lattice thermal conductivity is reduced drastically and the PF is improved by a few percent, the thermoelectric performance of $TaCu_3Te_4$ under 1 GPa pressure will be improved substantially.

The alloying through Se reduces the Seebeck coefficient largely while the electrical conductivity and electronic part of the thermal conductivity rise because the Se-substitution reduces the DOS near Fermi level and effective mass significantly. This type of change does not favor a high power factor, rather it reduces the PF significantly. On the other side, the lattice thermal conductivity is slightly reduced through alloying, suggesting that the alloying through the light element (such as Se) would not be favorable to improve TE performance of this family of compounds. For comparative and clarity purposes, the calculated transport coefficients and related parameters for n- and p-type carriers of all cases at 300K and 1000K are listed in Table IV.

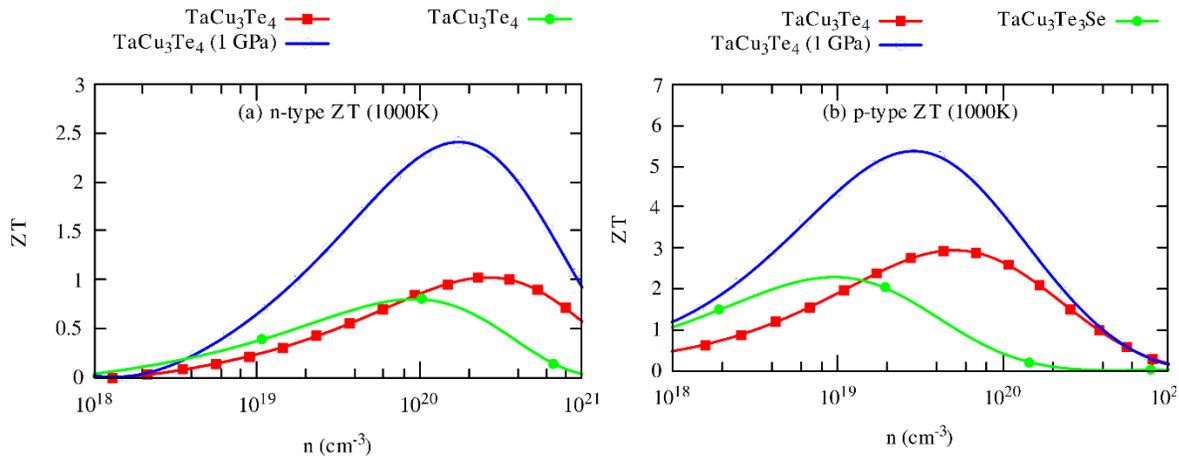

Fig. 9. Carrier concentration dependency of thermoelectric figure of merit (ZT) (at 1000K) of p-type $TaCu_3Te_4$ at ambient condition, under 1 GPa pressure, and alloyed through Se.

The 1 GPa pressure improves the ZT of $TaCu_3Te_4$ impressively, reaching a maximum value of ~5.5 at 1000K for p-type carriers, but alloying causes a substantial reduction of ZT, although lattice thermal conductivity is reduced slightly through the alloying. Such record-breaking ZT of a p-type bulk material, which has been already synthesized, will improve drastically the TE conversion efficiency. The ZT of both n- and p-type carriers is improved at 300K (see Fig. S15 in ESI), suggesting its potential applications at low and high temperatures. As the constituent elements are earth-abundant and non-toxic, future TE device may use this compound in high-temperature heat conversion management systems commercially.

4.  Conclusions

In summary, the effect of electron-phonon scattering, pressure, and alloying on the thermoelectric performance of $TmCu_3Ch_4$ has been studied by first-principles considering spin-orbit coupling effect, and energy-dependent carrier lifetime. The large discrepancies have been found between the present study and the reported thermoelectric performance of pristine $TmCu_3Ch_4$ and reasonable explanations have been provided. The lightest VCu$_3$S$_4$ has the highest value of lattice thermal conductivity and the heaviest TaCu$_3$Te$_4$ has the lowest value of it. The 1 GPa pressure has enhanced phonon scattering significantly compared to that by the substitution of Se. Thus, the κl drops to 0.17 from 0.38 W m$^{-1}$ K$^{-1}$ at 1000K. On the other side, all these compounds exhibit an indirect bandgap, with the values of gap 1.20-2.12 eV, and the heavier compound has a narrower gap and vice versa. The bandgap of TaCu$_3$Te$_4$ is slightly reduced by 1 GPa pressure (1.2 from 1.25 eV) and by Se-substitution (1.23 eV). However, the density of states near the Fermi level is reduced by Se-substitution significantly while the pressure does not cause any significant change. Interestingly, the power factor is slightly improved due to the increase of electrical conductivity by the slight reduction of bandgap under pressure. But the PF of Se-substituted alloy shows the opposite trend. Such type of changes of heat and carrier conduction mechanisms under pressure leads to an extraordinary ZT ~5.5 at 1000K for the p-type carrier, which is about two times larger than that of pristine TaCu$_3$Te$_4$ (~3 at 1000K). This is the record-breaking ZT reported to date for experimentally synthesized bulk material at such high temperatures.

## Data availability

All data are available upon reasonable request to the author.

Supplementary Information

Effect of electron-phonon scattering, pressure and alloying on the thermoelectric performance of TmCu$_3$Ch$_4$ (Tm=V, Nb, Ta; Ch=S, Se, Te)

Enamul Haque

EH Solid State Physics Laboratory, Longaer, Gaffargaon-2233, Mymensingh, Bangladesh.

Email: enamul.phy15@yahoo.com, enamul@mailaps.org

## S1. Computational Details

Table S1. Muffin tin sphere radii (Rmt) in Bohr.

| Compounds | | Rmt |
|---|---|---|
| VCu$_3$S$_4$ | V | 1.79 |
| | Cu | 1.91 |
| | S | 1.54 |
| VCu$_3$Se$_4$ | V | 1.78 |
| | Cu | 1.86 |
| | Se | 1.77 |
| VCu$_3$Te$_4$ | V | 1.90 |
| | Cu | 1.93 |
| | Te | 1.93 |
| NbCu$_3$S$_4$ | Nb | 1.87 |
| | Cu | 1.96 |
| | S | 1.61 |
| NbCu$_3$Se$_4$ | Nb | 1.85 |
| | Cu | 1.90 |
| | Se | 1.81 |
| NbCu$_3$Te$_4$ | Nb | 1.95 |
| | Cu | 1.97 |
| | Te | 1.97 |
| TaCu$_3$S$_4$ | Ta | 1.92 |
| | Cu | 1.97 |
| | S | 1.57 |
| TaCu$_3$Se$_4$ | Ta | 1.89 |
| | Cu | 1.91 |
| | Se | 1.80 |
| TaCu$_3$Te$_4$ | Ta | 2.00 |
| | Cu | 1.97 |
| | Te | 1.97 |
| TaCu$_3$Te$_4$ (1 GPa) | Ta | 2.00 |
| | Cu | 1.96 |
| | Te | 1.96 |
| TaCu$_3$Te$_3$Se | Ta | 1.75 |
| | Cu | 1.77 |
| | Te | 1.82 |
| | Se | 1.66 |

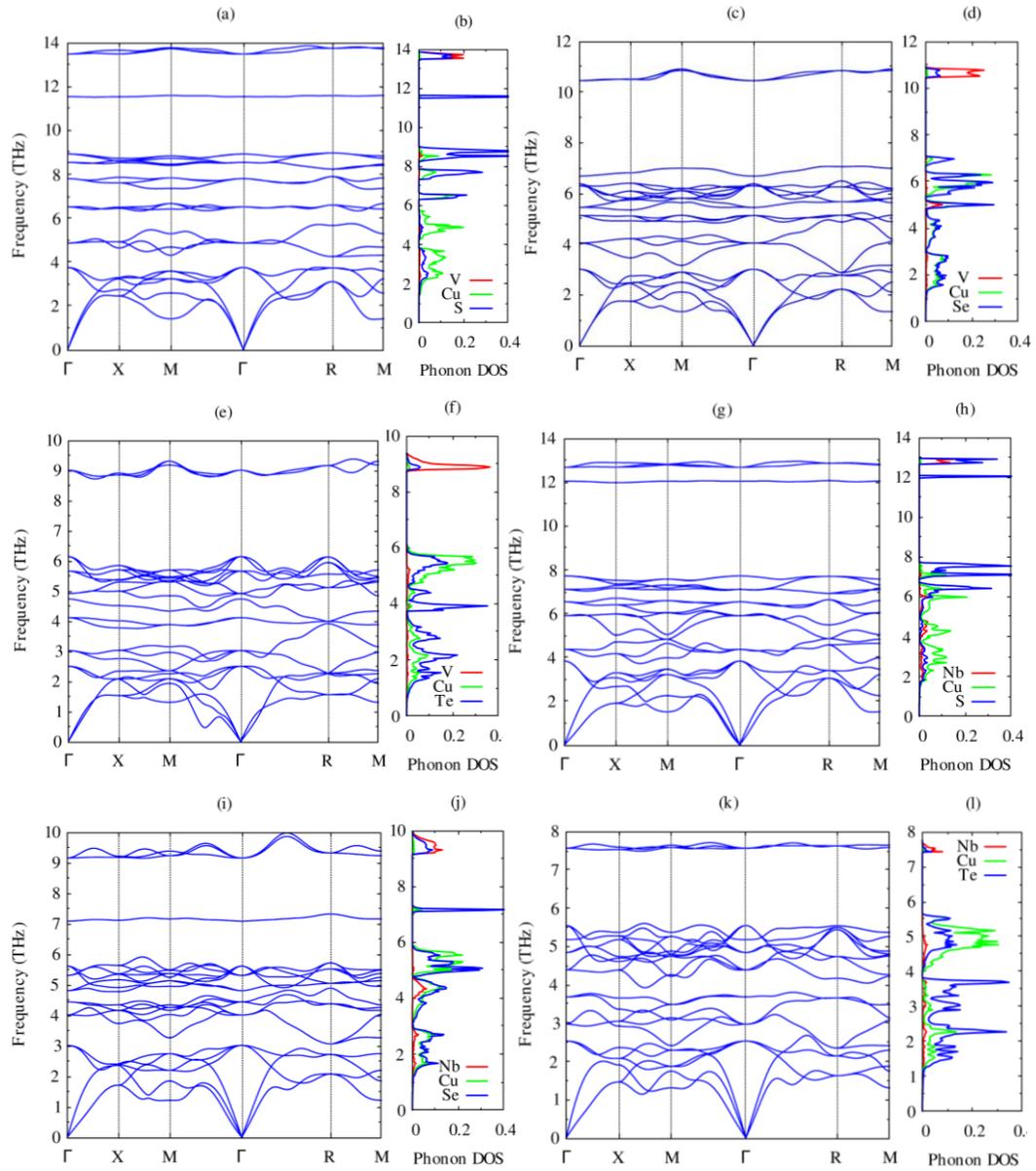

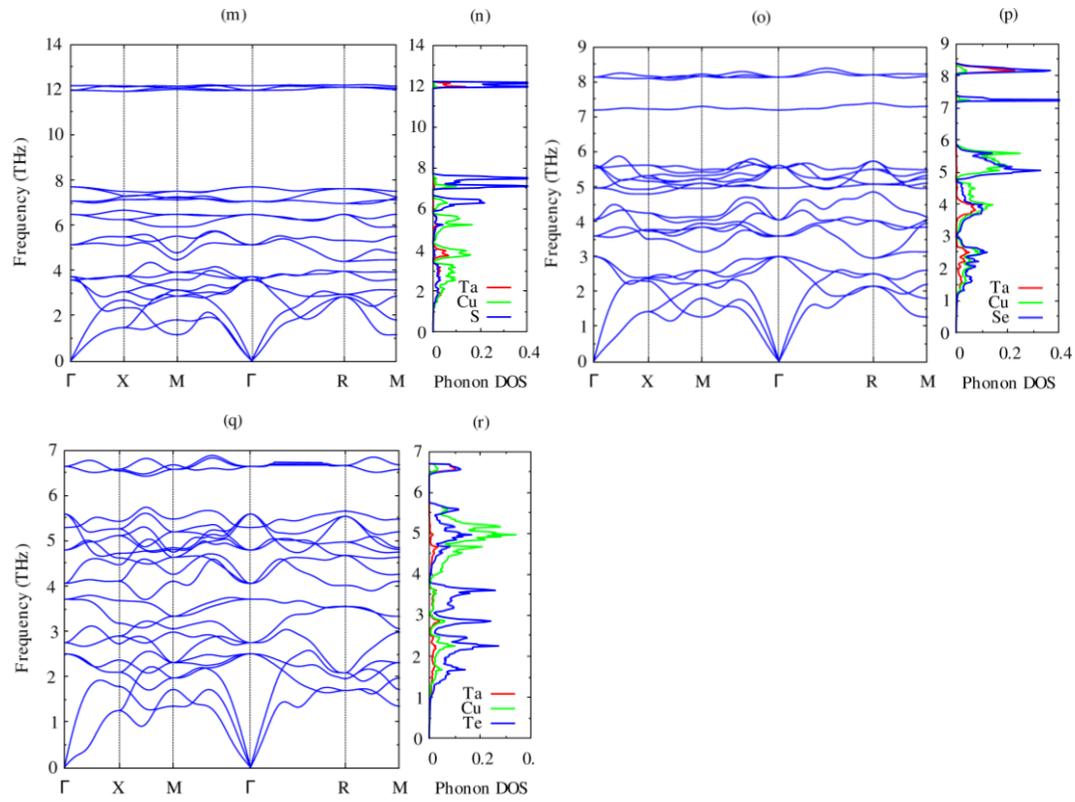

Fig. S1. Phonon dispersion relations and partial phonon density of states of pristine: (a-b) $VCu_3S_4$, (c-d) $VCu_3Se_4$, (e-f) $VCu_3Te_4$, (g-h) $NbCu_3S_4$, (i-j) $NbCu_3Se_4$, (k-l) $NbCu_3Te_4$, (m-n) $TaCu_3S_4$, (o-p) $TaCu_3Se_4$, (q-r) $TaCu_3Te_4$.

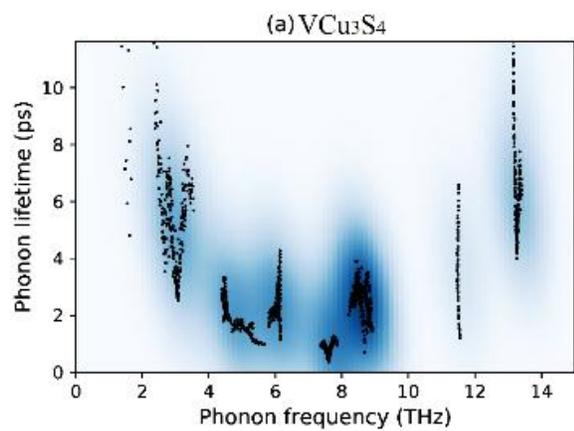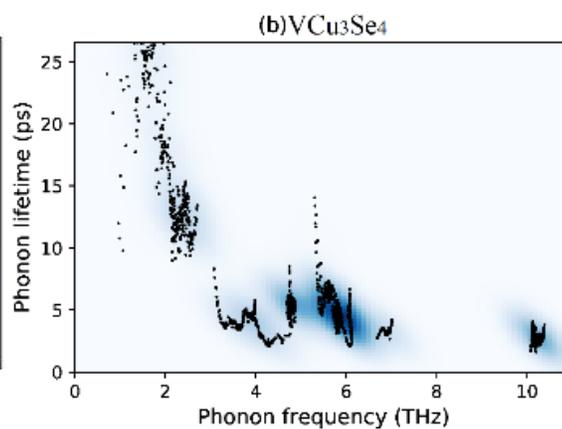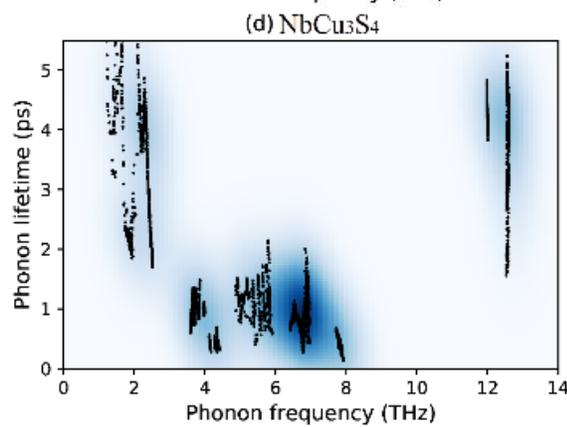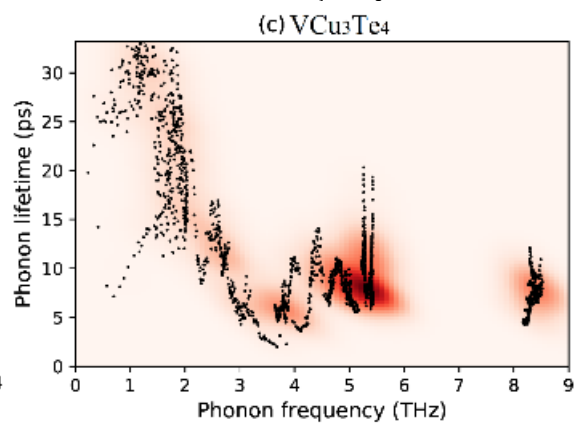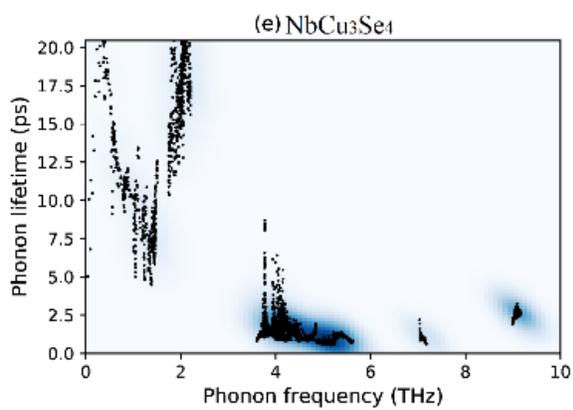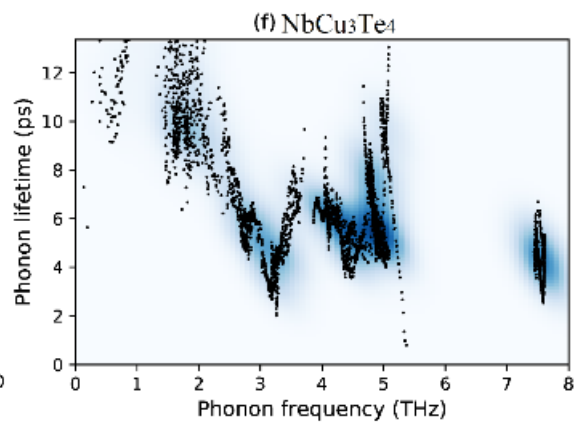

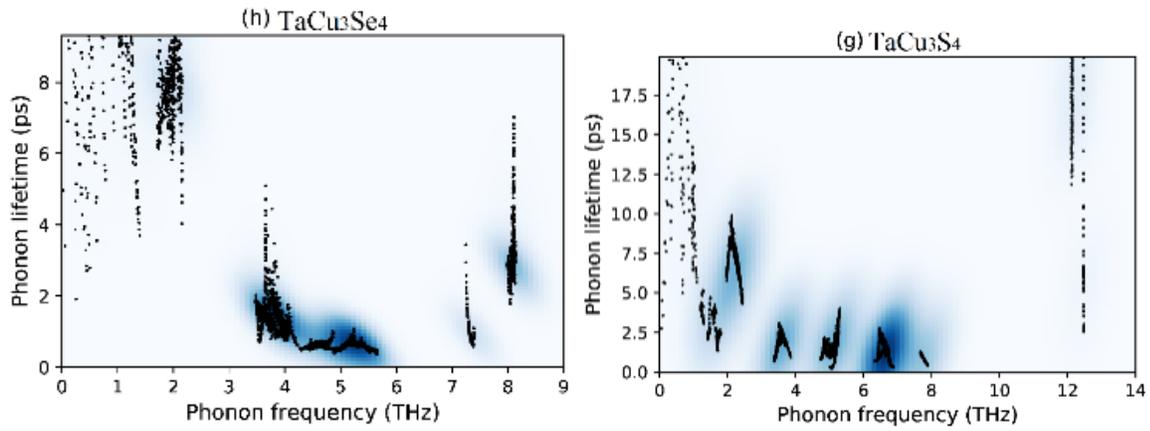

Fig. S2. Calculated phonon lifetime from second and third order IFCs of pristine: (a) VCu$_3$S$_4$, (b) VCu$_3$Se$_4$, (c) VCu$_3$Te$_4$, (d) NbCu$_3$S$_4$, (e) NbCu$_3$Se$_4$, (f) NbCu$_3$Te$_4$, (g) TaCu$_3$S$_4$, and (h) TaCu3Se4.

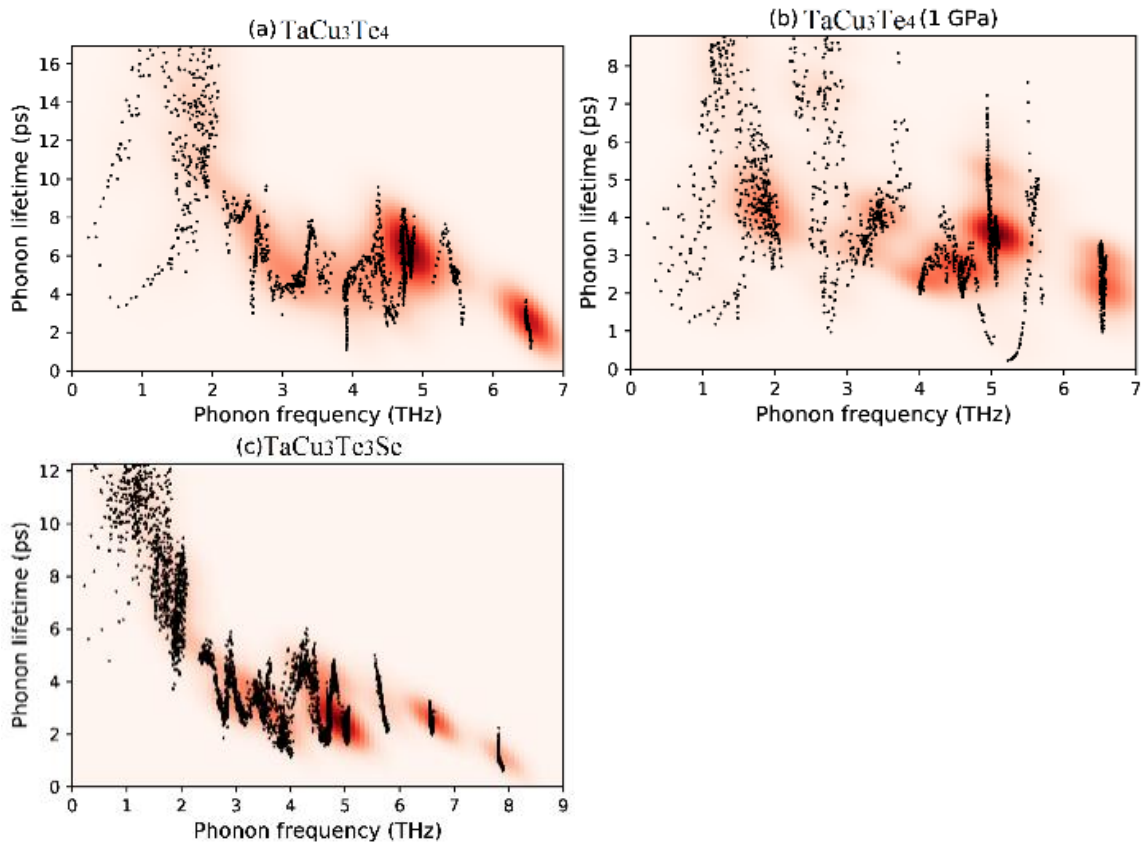

Fig. S3. Calculated phonon lifetime from second and third order IFCs of: (a) TaCu$_3$Te$_4$, (b) TaCu3Te4 (1 GPa), and (c) TaCu$_3$Te$_3$Se

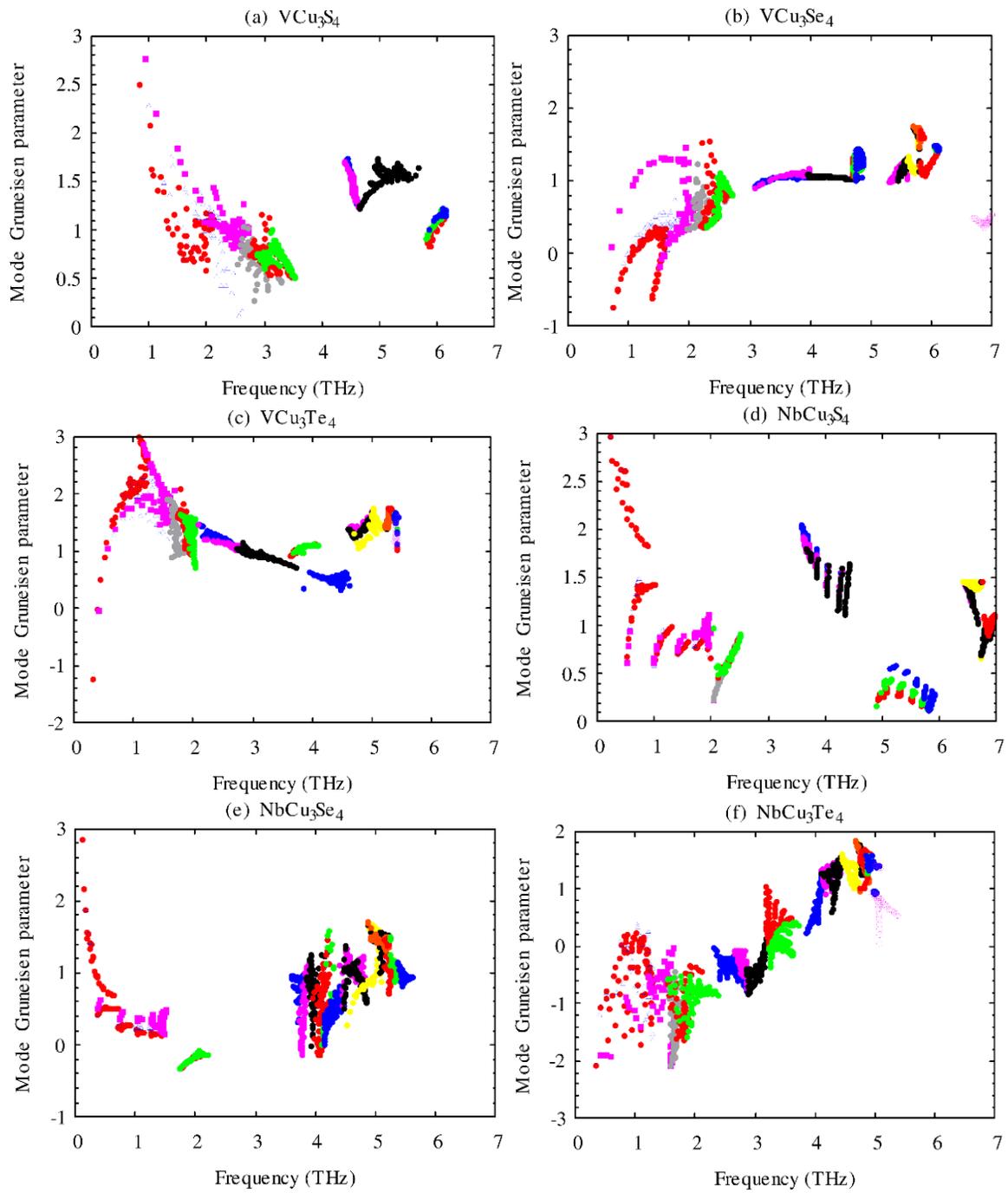

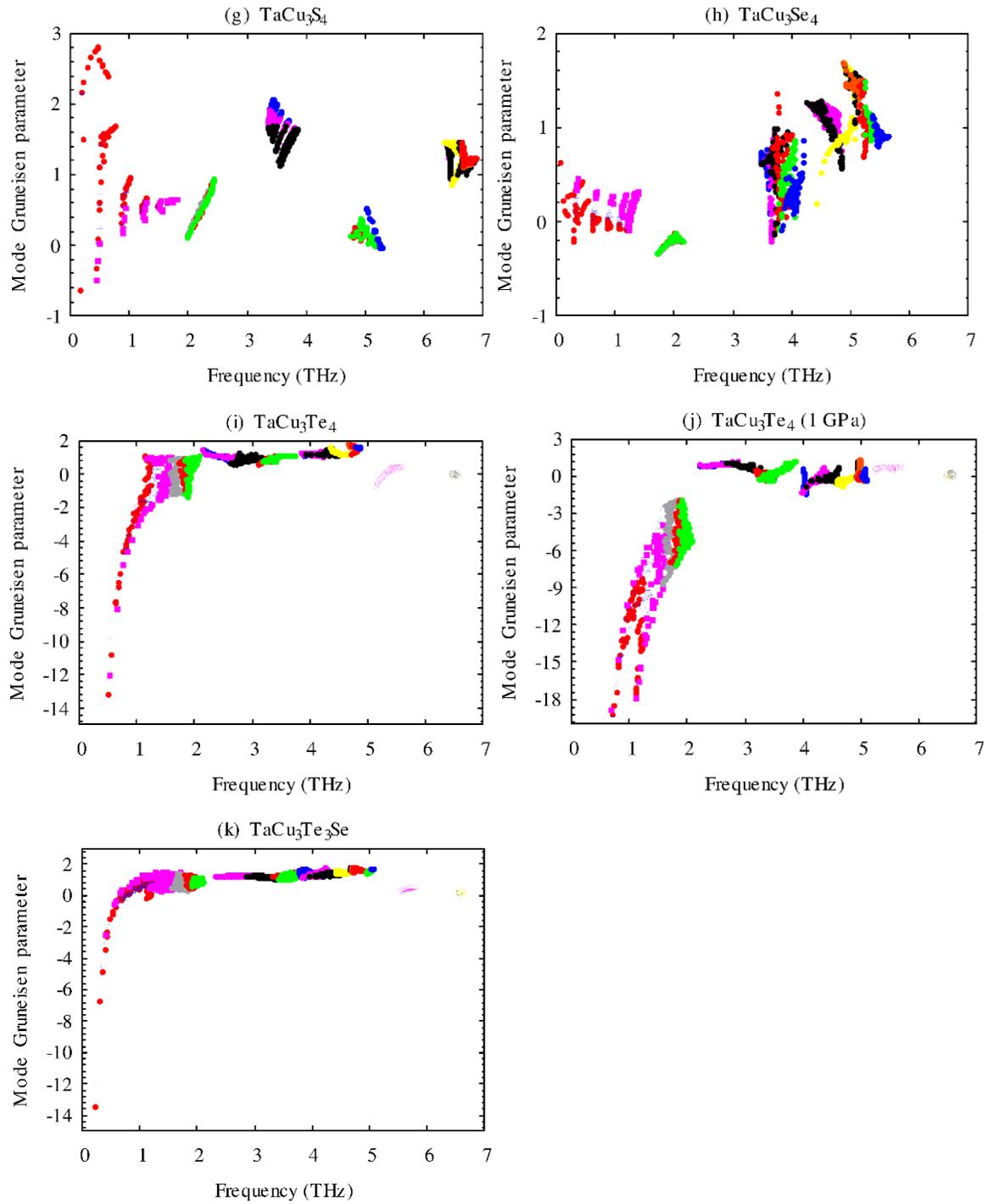

Fig. S4. Computed mode Gruneisen parameter from second and third order IFCs of pristine: (a) $VCu_3S_4$, (b) $VCu_3Se_4$, (c) $VCu_3Te_4$, (d) $NbCu_3S_4$, (e) $NbCu_3Se_4$, (f) $NbCu_3Te_4$, (g) $TaCu_3S_4$, (h) TaCu3Se4, (i) $TaCu_3Te_4$, (j) TaCu3Te4 (1 GPa), and (k) $TaCu_3Te_3Se$.

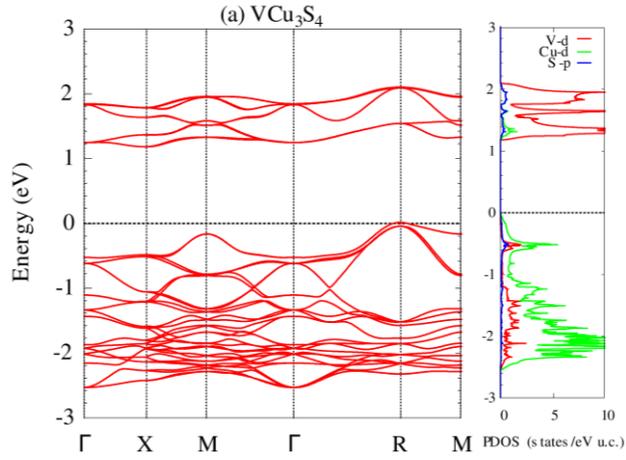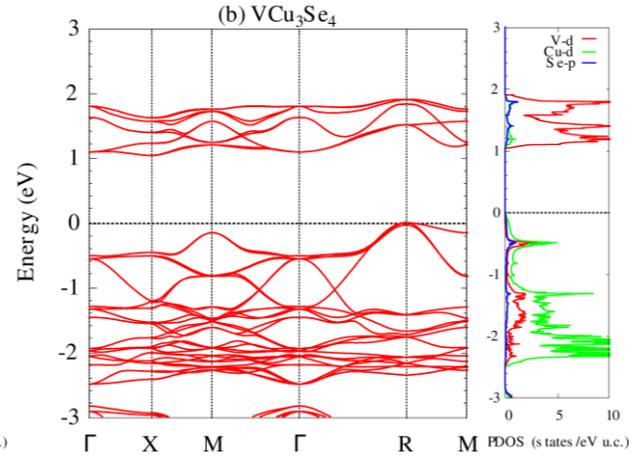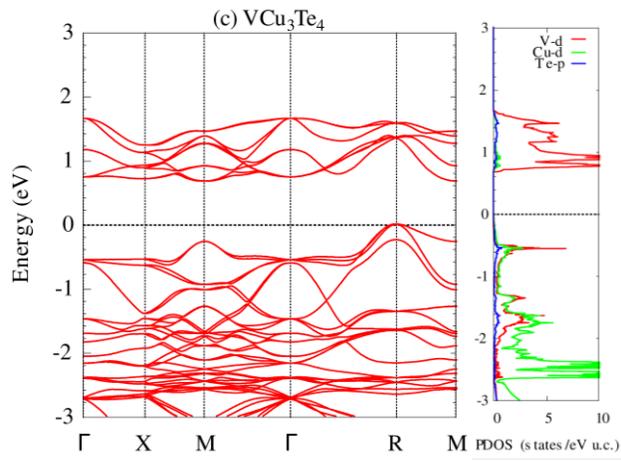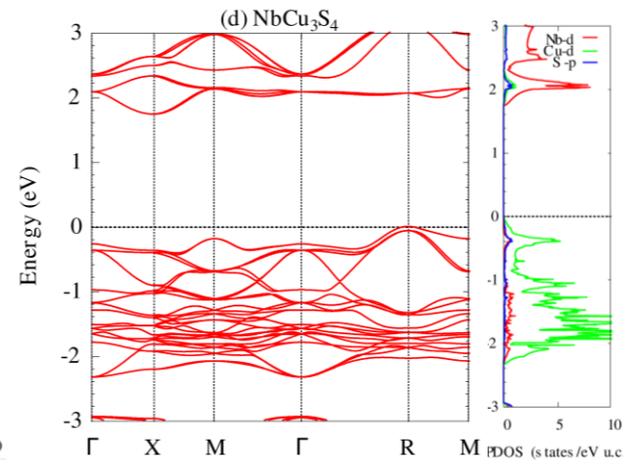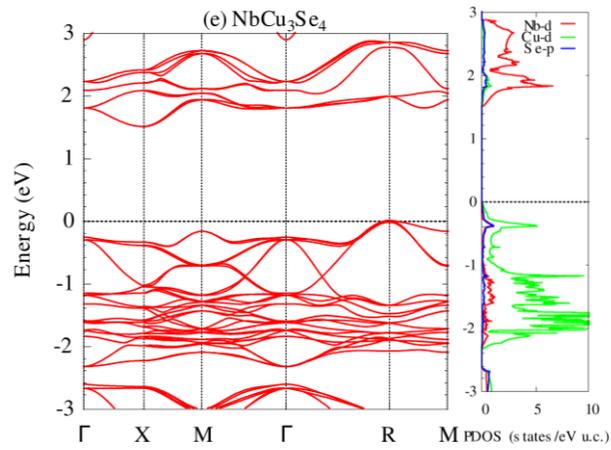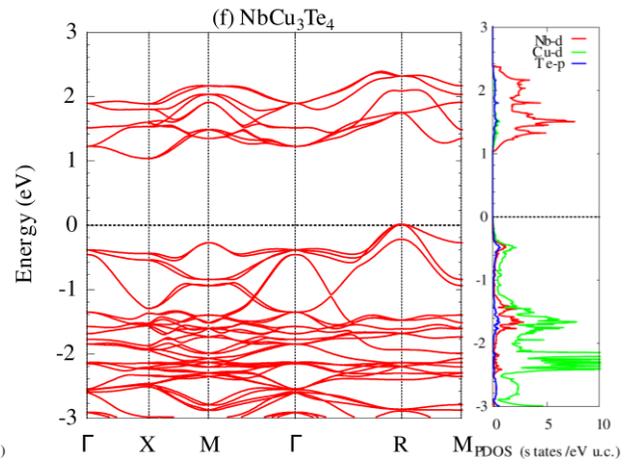

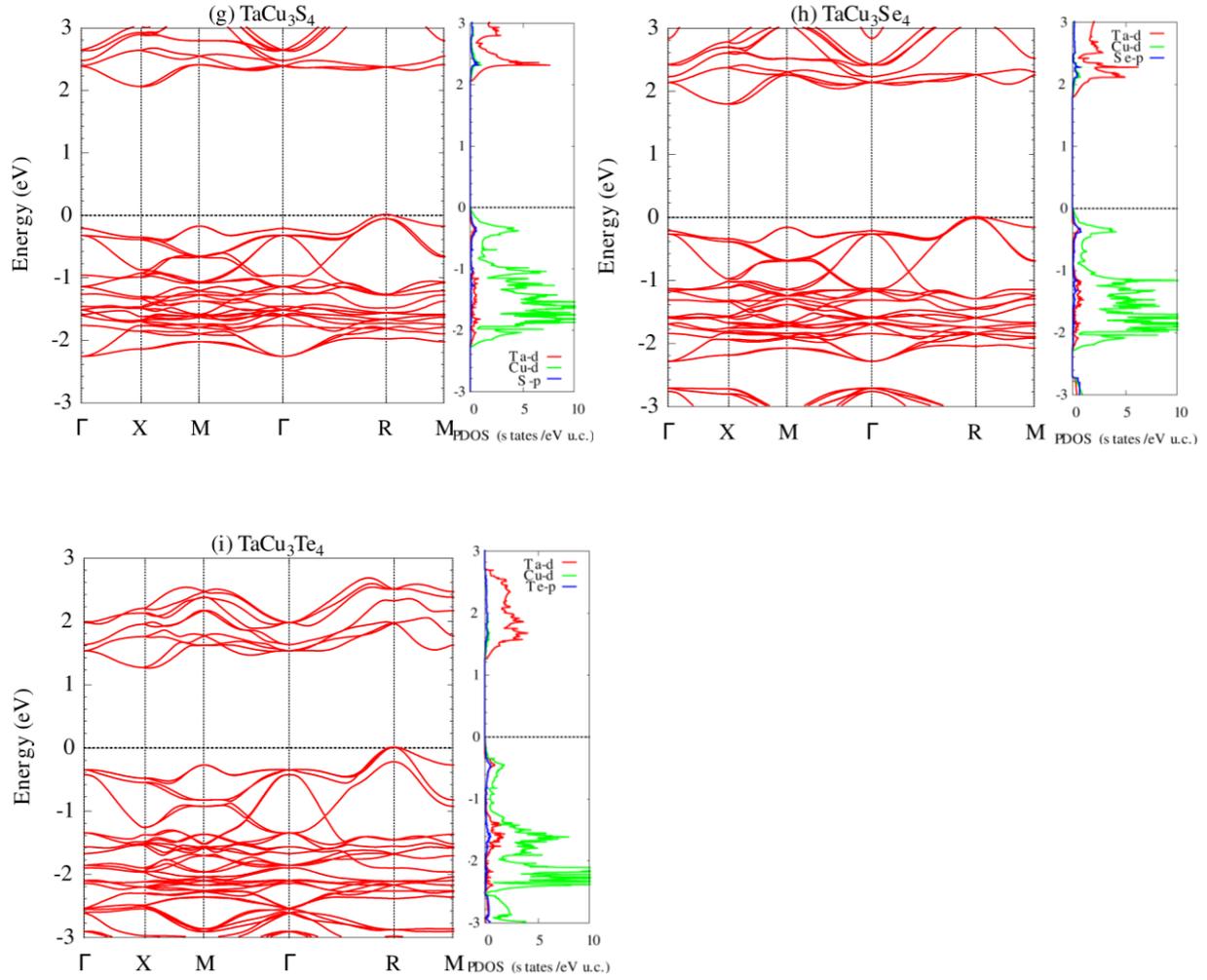

Fig. S5. Computed electronic band structures and projected density of states, by using mBJ potential including spin-orbit coupling effect, of: (a) $VCu_3S_4$, (b) $VCu_3Se_4$, (c) $VCu_3Te_4$, (d) $NbCu_3S_4$, (e) $NbCu_3Se_4$, (f) $NbCu_3Te_4$, (g) $TaCu_3S_4$, (h) $TaCu3Se4$, and (i) $TaCu_3Te_4$. The right panel of each sub-figure presents the projected density of states and the grey colored dashed line at zero energy is the Fermi level.

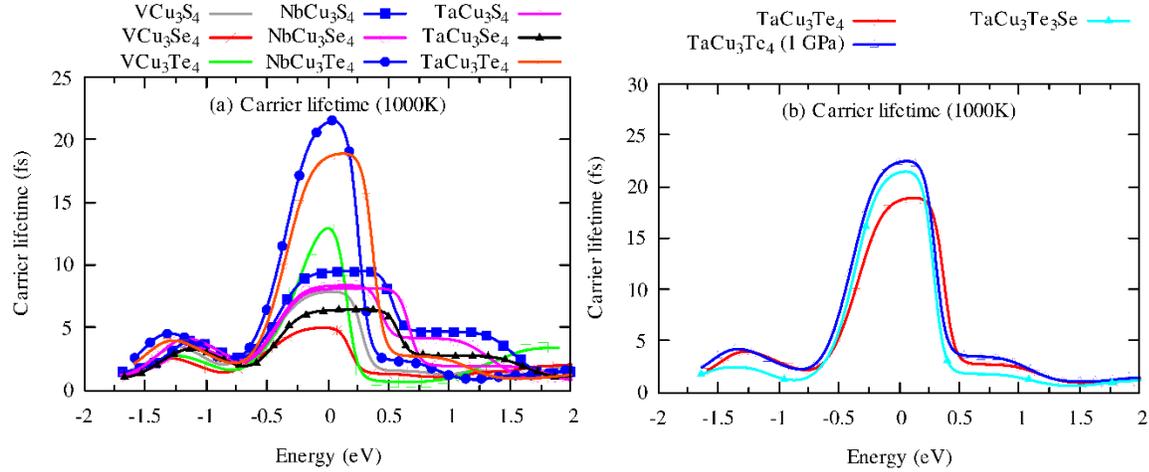

Fig. S6. Computed energy-dependent carrier lifetime of TmCu$_3$Ch$_4$ (Tm=V, Nb, Ta; Ch=S, Se, Te), TaCu$_3$Te$_4$ under I GPa, and TaCu$_3$Te$_3$Se at 1000K.

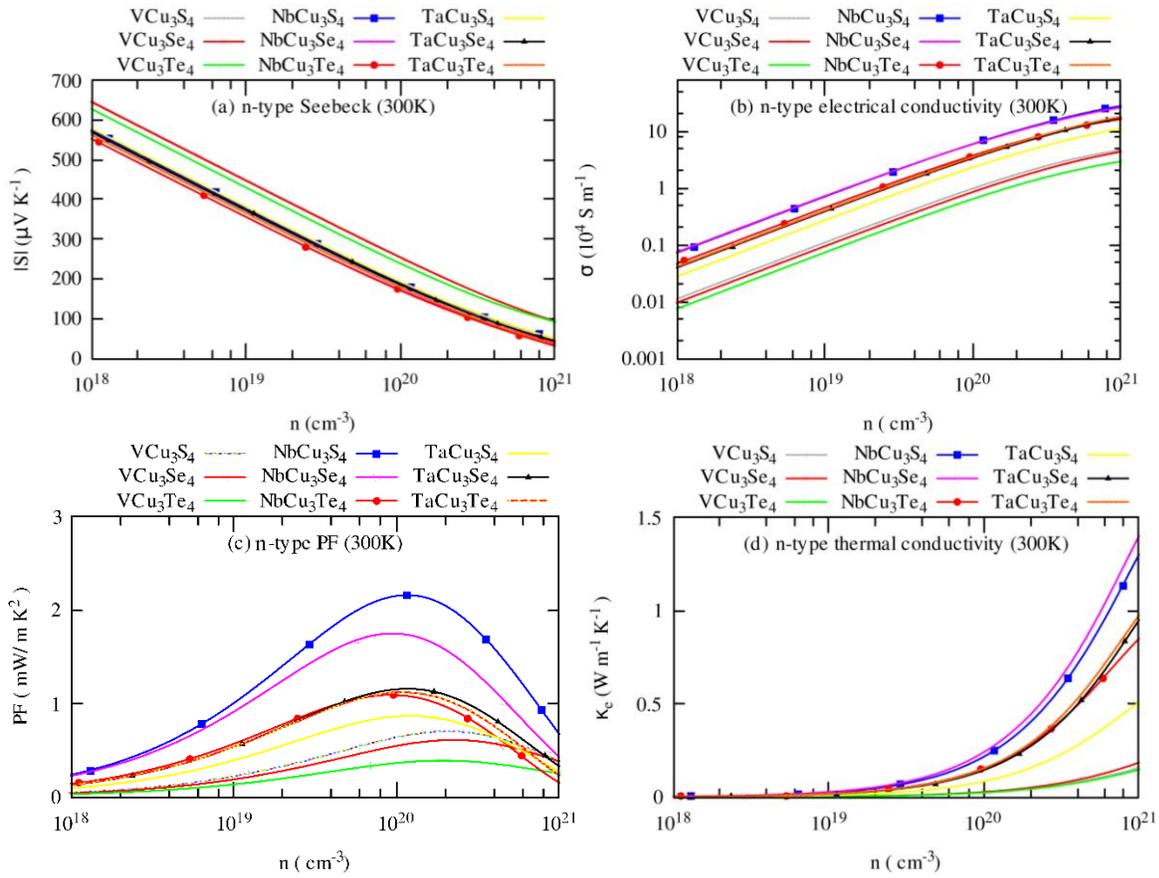

Fig. S7. Computed transport coefficients as a function of carrier-concentration of n-type TmCu$_3$Ch$_4$ (Tm=V, Nb, Ta; Ch=S, Se, Te) at 300K: (a) Seebeck coefficient, (b) electrical conductivity, (c) power factor (PF), and (a) electronic part of the thermal conductivity ($\kappa_e$).

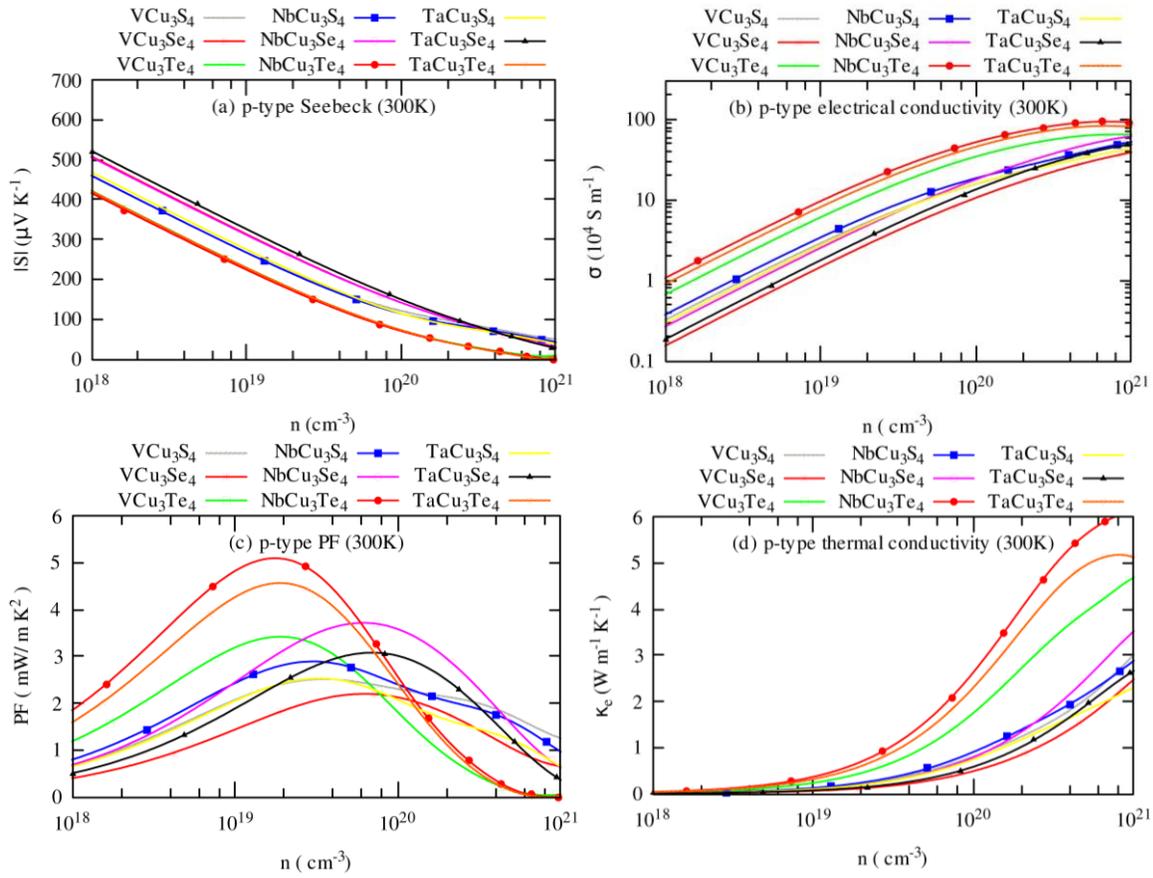

Fig. S8. Computed transport coefficients as a function of carrier-concentration of p-type TmCu$_3$Ch$_4$ (Tm=V, Nb, Ta; Ch=S, Se, Te) at 300K: (a) Seebeck coefficient, (b) electrical conductivity, (c) power factor (PF), and (a) electronic part of the thermal conductivity ($\kappa_e$).

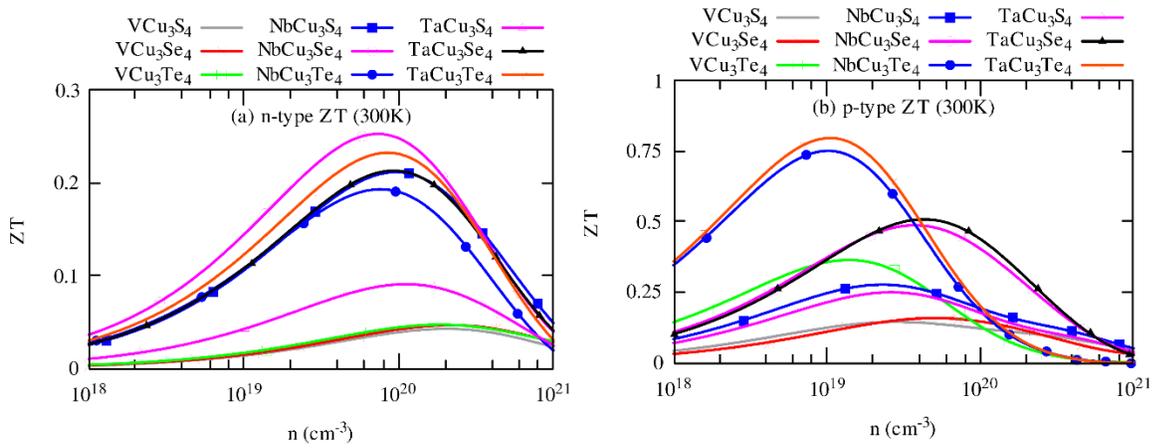

Fig. S9. Computed room temperature thermoelectric figure of merit as a function of carrier concentration of TmCu$_3$Ch$_4$ (Tm=V, Nb, Ta; Ch=S, Se, Te) for (a) n-type and (b) p-type carriers.

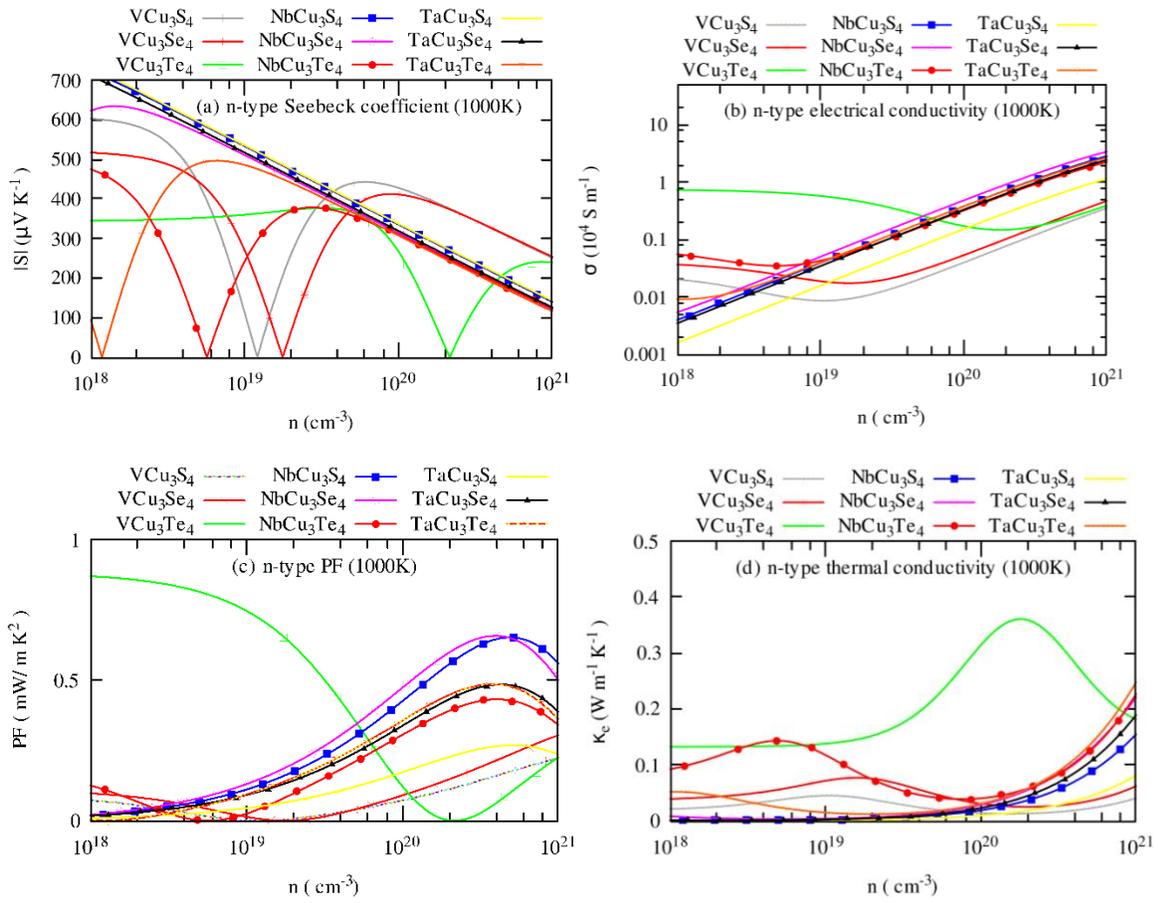

Fig. S10. Computed transport coefficients as a function of carrier-concentration of n-type TmCu$_3$Ch$_4$ (Tm=V, Nb, Ta; Ch=S, Se, Te) at 1000K: (a) Seebeck coefficient, (b) electrical conductivity, (c) power factor (PF), and (a) electronic part of the thermal conductivity ($\kappa_e$).

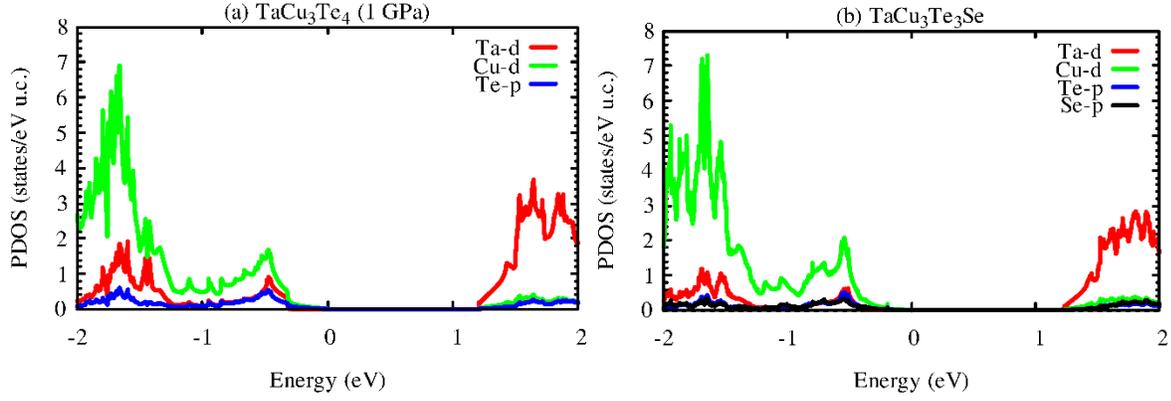

Fig. S11. Projected density of states of TaCu3Te4 under 1 GPa pressure and TaCu3TeSe. The zero-energy represents the Fermi level.

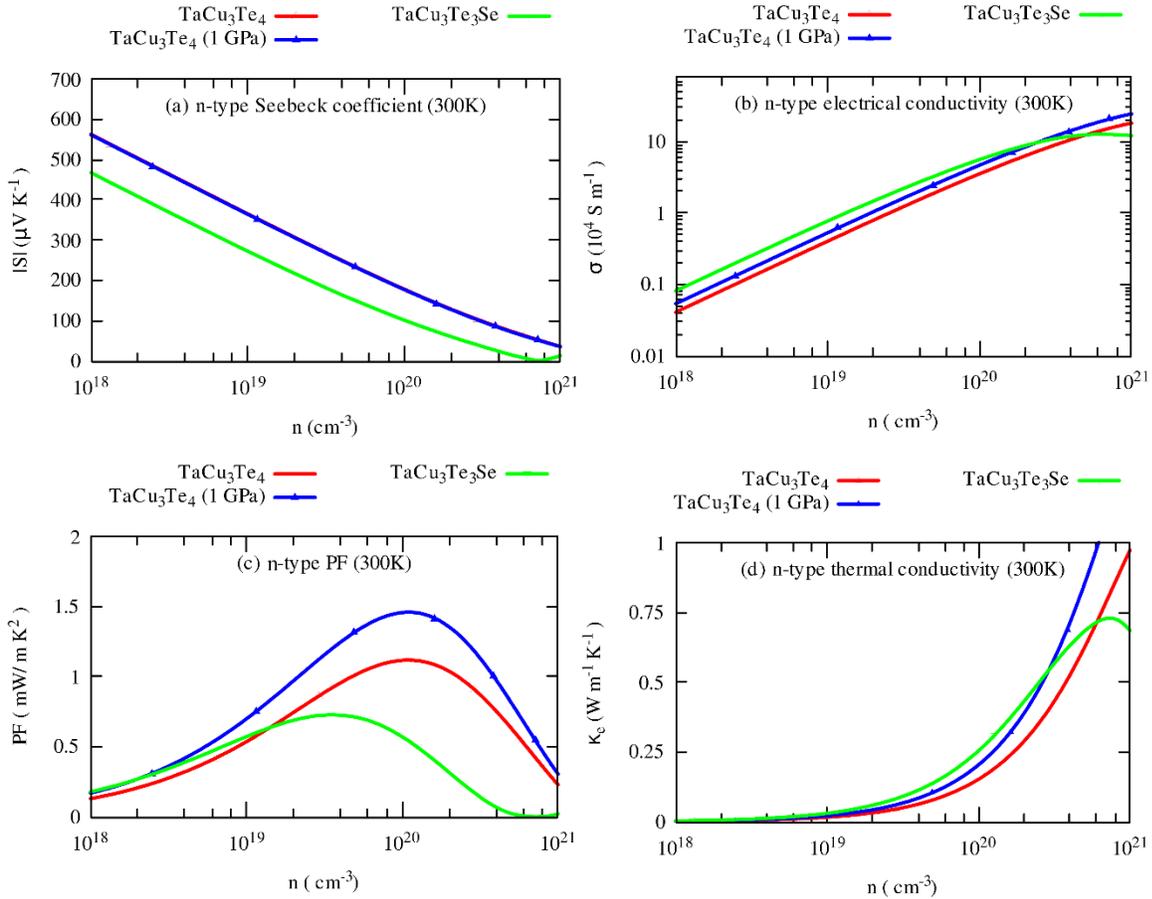

Fig. S12. Carrier concentration-dependent transport coefficients (at 300K) of n-type $TaCu_3Te_4$ at ambient condition, under 1 GPa pressure, and $TaCu_3Te_3Se$.

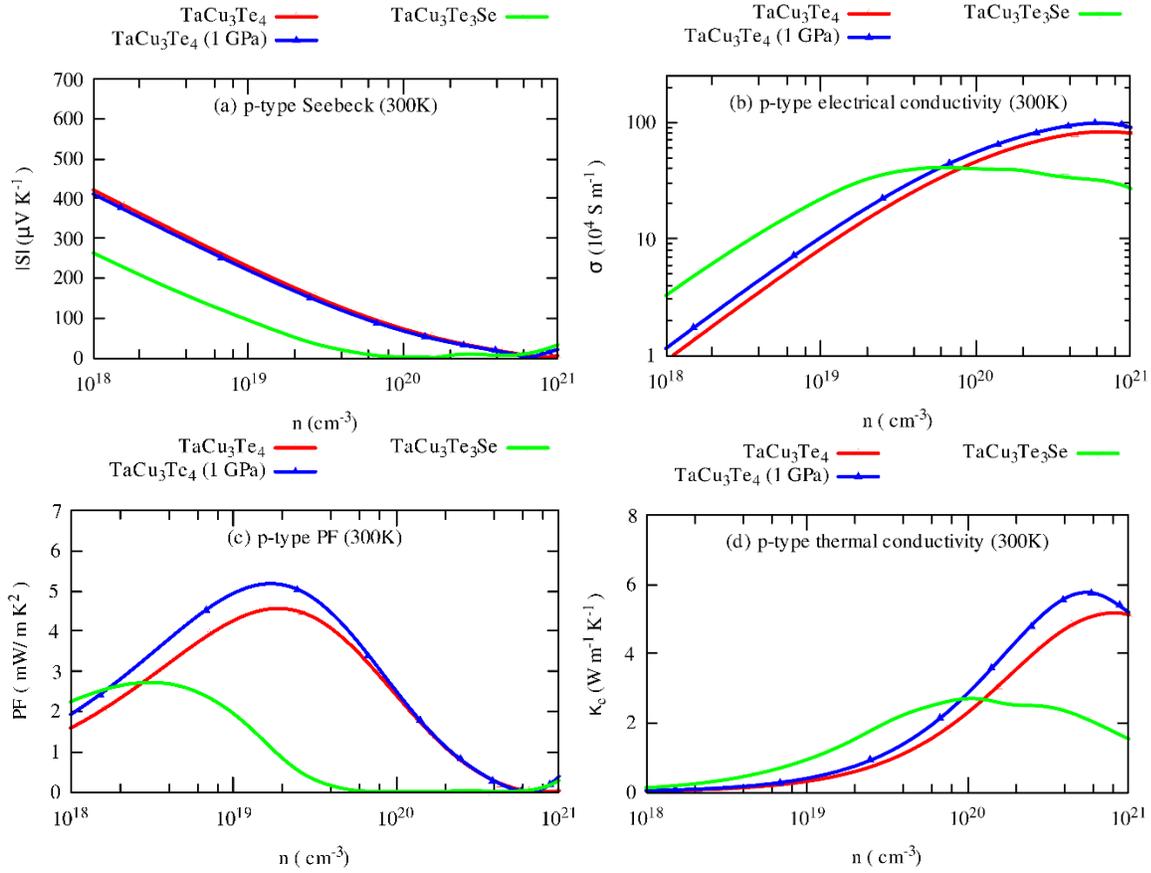

Fig. S13. Carrier concentration-dependent transport coefficients (at 300K) of p-type TaCu$_3$Te$_4$ at ambient condition, under 1 GPa pressure, and TaCu$_3$Te$_3$Se.

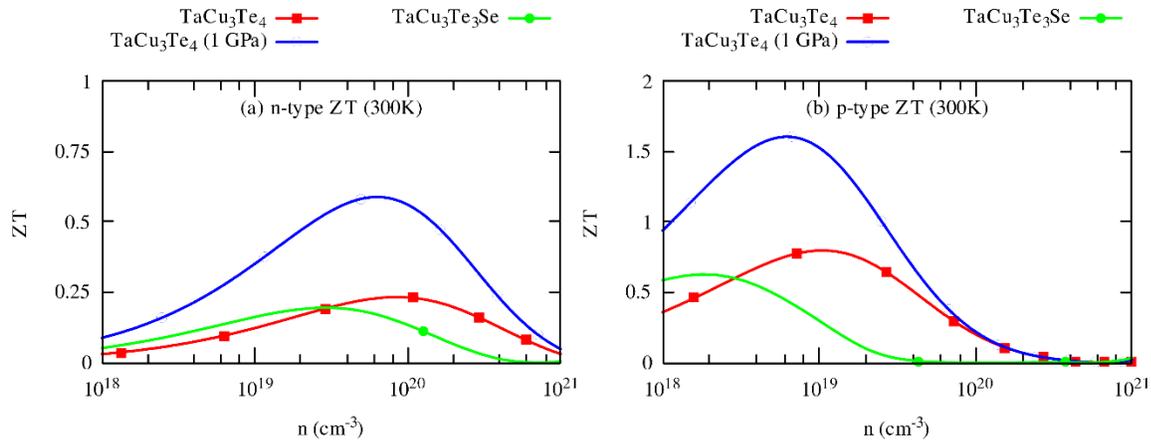

Fig. S14. Calculated thermoelectric figure of merit (at 300K) of TaCu$_3$Te$_4$ at ambient condition, under 1 GPa pressure, and TaCu$_3$Te$_3$Se for (a) n-type and (b) p-type carriers.

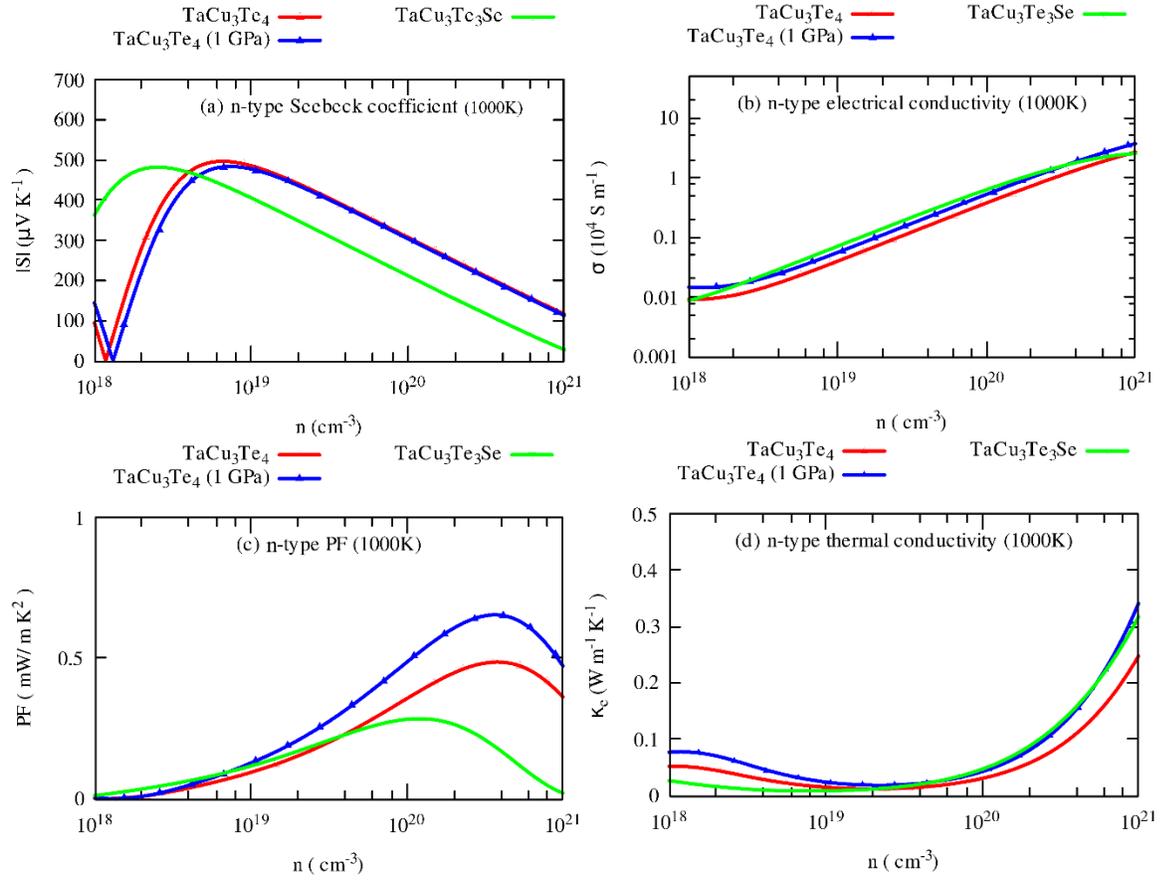

Fig. S15. Computed transport coefficients as a function of carrier density at 1000K of n-type TaCu$_3$Te$_4$ at ambient condition, under 1 GPa pressure, and TaCu$_3$Te$_3$Se.